\documentclass[12pt, a4paper]{article}
\pdfoutput=1
\usepackage{jheppub_mod}
\usepackage{amsmath,amssymb}
\usepackage{graphicx} 
\usepackage{comment}
\usepackage{multirow}
\usepackage{setspace}
\usepackage{color}
\allowdisplaybreaks
 \title{Weights and Recursion relations for $\phi^p$ Tree Amplitudes from the Positive Geometry}
 \author{Ryota Kojima}
\affiliation{KEK Theory Center, Tsukuba, Ibaraki, 305-0801, Japan}
\emailAdd{ryota@post.kek.jp}
\abstract{
Recently, the accordiohedron in kinematic space was proposed as the positive geometry for planar tree-level scattering amplitudes in the $\phi^p$ theory \cite{Raman:2019utu}. The scattering amplitudes are given as a weighted sum over canonical forms of some accordiohedra with appropriate weights. These weights were determined by demanding that the weighted sum corresponds to the scattering amplitudes. It means that we need additional data from the quantum field theory to compute amplitudes from the geometry. It has been an important problem whether scattering amplitudes are completely obtained from only the geometry even in this $\phi^p$ theory.\par
In this paper, we show that these weights are completely determined by the factorization property of the accordiohedron. It means that the geometry of the accordiohedron is enough to determine these weights. 
\par
In addition to this, we study one-parameter recursion relations for the $\phi^p$ amplitudes. The one-parameter ``BCFW"-like recursion relation for the $\phi^3$ amplitudes was obtained from the triangulation of the ABHY-associahedron \cite{Arkani-Hamed:2017tmz}. After this, a new recursion relation was proposed from the projecting triangulation of the generalized ABHY-associahedron in \cite{Arkani-Hamed:2019vag, Yang:2019esm}. We generalize these one-parameter recursion relations to the $\phi^p$ amplitudes and interpret as triangulations of the accordiohedra. 
}
  \allowdisplaybreaks[1]
\begin{document}
%-------------------------------------------------------------------------

\begin{flushright}
KEK-TH-2212
\end{flushright}

\maketitle
\newpage
\section{Introduction}
Recent years have revealed an unexpected connection between scattering amplitudes and the positive geometry \cite{Arkani-Hamed:2013jha, Arkani-Hamed:2013kca, Arkani-Hamed:2017vfh,Arkani-Hamed:2017tmz, Arkani-Hamed:2017fdk, Arkani-Hamed:2018ign,Arkani-Hamed:2019mrd,Arkani-Hamed:2019vag,Arkani-Hamed:2019rds}. This connection gives a purely geometric definition of scattering amplitudes. In particular, tree amplitudes for massless bi-adjoint $\phi^3$ theory are given as the canonical form of an ABHY-associahedron defined in kinematic space \cite{Arkani-Hamed:2017mur}. This geometric picture is generalized for more general interaction cases such as the stokes polytope for $\phi^4$ theory \cite{Banerjee:2018tun}, the accordiohedron for $\phi^p$ theory \cite{Raman:2019utu}, and the case of the polynomial interaction \cite{Jagadale:2019byr}. In the $\phi^p (p>3)$ case, the geometry is a union of many polytopes of a given dimension. The canonical form of each polytope gives a partial amplitude which is given from the subsets of $\phi^p$ graphs. We need to sum over all canonical forms for each polytope with appropriate weight to obtain tree-level planar amplitudes. These weights are determined uniquely from the condition that when summed over all canonical forms, the residue of each pole is unit. Once we choose these weights from this condition, the sum of canonical forms corresponds to the scattering amplitude. This condition is obtained by demanding that the weighted sum corresponds to the scattering amplitudes \cite{Raman:2019utu}.
The important point is that the derivation of this condition relied on the form of scattering amplitudes. It has been an important problem whether these weights are completely determined from the geometry only even in this $\phi^p$ theory.\par
This geometric picture of the amplitudes gives us a new viewpoint of the recursion relations. In \cite{Arkani-Hamed:2017mur}, a one-parameter recursion relation for tree bi-adjoint $\phi^3$ amplitudes is derived from the triangulation of the ABHY-associahedron. This relation of the recursion relation and triangulation is similar to the relation between ``BCFW recursion relation" \cite{Britto:2004ap, Britto:2005fq} of the tree NMHV amplitude in the $\mathcal{N}=4$ SYM and the triangulation of the amplituhedron \cite{Arkani-Hamed:2013jha, Arkani-Hamed:2013kca}. From this ``BCFW-like" recursion for $\phi^3$ amplitudes, an all-multiplicity results is obtained in \cite{He:2018svj}. Recently, a new recursion relation was proposed from the projecting triangulation of the generalized ABHY-associahedron \cite{Arkani-Hamed:2019vag}. This ``projective recursion" was derived from field-theoretical consideration in \cite{Yang:2019esm} and it can be interpreted as a generalization of the BCFW-like recursion. It is known that the special case of this projective recursion: ``one-variable recursion" is much more efficient than the BCFW-like recursion \cite{Yang:2019esm}. This recursion can be also derived from the general properties of canonical forms of simple polytopes \cite{Salvatori:2019phs}.\par

In this paper, we investigate the weights of the accordiohedra and one-parameter recursion relations of the $\phi^p$ amplitudes. We show that these weights are completely determined from the factorization property of the accordiohedron. This means that the geometry of the accordiohedron is enough to determine these weights. We can also see that the condition that each pole has unit residue is equivalent to one of the constraints obtained from the factorization. \par
We generalize the one-parameter recursions for $\phi^p$ amplitudes. The important point is that the recursion relation can be defined to the subset of $\phi^p$ graphs, not to all of the graphs. This comes from the fact that in the $\phi^p (p>3)$ case, the scattering form on the kinematic space can be defined uniquely from the projectivity only for the subset of $\phi^p$ graphs. This projectivity means that partial amplitudes have no pole at infinity in the kinematic space. We need this property to construct the recursion relation. 
%First, we consider the BCFW-like recursion for $\phi^p$ amplitudes. We show that this recursion can be interpreted as a triangulation of the accordiohedra. By using the general formula of the triangulation, we obtain an all-multiplicity result. Next, we attempt to apply the projective recursion for $\phi^p$ amplitudes. We show some explicit calculations of the one-variable recursion. We will see that this recursion is much more efficient than the BCFW-like recursion.
\par
This paper is organized as follows: in section \ref{sec:review}, we briefly review the positive geometry of the $\phi^p$ tree amplitude. In section \ref{sec:weightfactorization}, we will see how to determine all the weights from the factorization property of the accordiohedron. In section \ref{sec:recursion relation}, we will construct the BCFW-like recursion relation for $\phi^p$ amplitudes. The  recursion relation is applied to each subset of graphs. We also see that this recursion can be interpreted as a triangulation of the accordiohedra and the general formula of the triangulation leads all-multiplicity results. Finally, in section \ref{sec:projectiverecursion}, we will construct the projective recursion relation for $\phi^p$ amplitudes. We show some explicit calculations of the one-variable projective recursion which is much more efficient than the BCFW-like recursion.
\section{Positive Geometry for $\phi^p$ tree amplitudes}
\label{sec:review}
%-----------------------------------------------------------------------------------------------------------
\subsection{Accordiohedron}
Here we define a geometric object called the ``accordiohedron" \cite{alex2016oriented}, which is proposed as the positive geometry for $\phi^p (p>3)$ interactions \cite{Raman:2019utu}. First we introduce the notion of ``$Q$-compatible diagonal" and ``$Q$-flip".  Let us consider the decomposition of convex polygon into $p$-gons and we call this decomposition as $p$-angulation. We can $p$-angulate a $(2p-2)$-gon into two $p$-gons and there are $p-1$ $p$-angulations. We denote these $p$-angulations by using the diagonals $\{(1,p),(2,p+1),\dots,(p-1,2p-2)\}$. We introduce a notion of ``$Q$-compatible diagonal" to each diagonal $(i,j)$ of a $(2p-2)$-gon as:
\begin{equation}
\label{eq:Qcompatiblerule}
(i,j)\rightarrow(Mod(i+p-2,2p-2),Mod(j+p-2,2p-2)).
\end{equation}  
This can be generalized to any $p$-angulation of an $n$-gon. To do this, we need to consider the unique $(2p-2)$-gon which contains the diagonal $(i,j)$. Let's consider $p=4,n=8$ case and the quadrangulation $\{(1,4),(1,6)\}$ as an example. The unique hexagon which includes the diagonal $(1,6)$ is $\{1,4,5,6,7,8\}$ and we label this as $\{1',2',3',4',5',6'\}$. Then the $Q$-compatible diagonal of $(1,6)$ is given as 
\begin{equation}
(Mod(3,6),Mod(8,6))\rightarrow (3',6')=(5,8).
\end{equation}  
The operation of replacing a diagonal with its $Q$-compatible diagonal is called ``$Q$-flip". \par
Next, we define the accordiohedron $\mathcal{AC}^{P}_{p,n}$ by using this $Q$-flip as follows.
When we $p$-angulate an $n$-gon into $C_p=(n-2)/(p-2)$ cells with $D_p=(n-p)/(p-2)$ diagonals, we call this as complete $p$-angulation\footnote{ Only $p+m(p-2)$-point polygon ($m=0,1,\dots$) can be $p$-angulated completely. If we substitute $p+m(p-2)$ for $n$ ,we can easily verify that these $C_p, D_p$ are integers.}. First we choose a complete $p$-angulation $P$ of the $n$-gon. For each $D_p$ diagonals of $P$, we apply the $Q$-flip operation until we do not generate any new $p$-angulation, and we obtain a subset of the $p$-angulations. We call this subset as $Q$-compatible set of $P$ ``$Q(P)$". Then the accordiohedron $\mathcal{AC}^{P}_{p,n}$ for this $P$ is defined as:
\begin{center}
Vertices $\leftrightarrow$ $p$-angulations in the $Q(P)$\\
Edges $\leftrightarrow$  $Q$-Flips\\
$k$-Facets $\leftrightarrow$ $k$-partial $p$-angulations\\
\end{center}
We can easily verify that $\mathcal{AC}^{P}_{p,n}$ depends on the $p$-angulation $P$ and the number of vertices of $n$-gon.
\par
The accordiohedron contains both associahedron and stokes polytope as special cases. In the case of $p=3$,  \eqref{eq:Qcompatiblerule} reduces to $(i,j) \rightarrow (Mod(i+1,4),Mod(j+1,4))$ and this is the mutation rule of the triangulation \cite{Arkani-Hamed:2017mur}. Then the $\mathcal{AC}^{P}_{3,n}$ corresponds to the associahedron. In the case of $p=4$,  \eqref{eq:Qcompatiblerule} reduces to $(i,j) \rightarrow (Mod(i+2,6),Mod(j+2,6))$ and this is the $Q$-compatibility of the quadrangulation \cite{Banerjee:2018tun}. From this, $\mathcal{AC}^{P}_{4,n}$ corresponds to the stokes polytope. 

%-----------------------------------------------------------------------------------------------------------
\subsection{Planar kinematic variables and the scattering form}
Kinematic space of massless momenta $p_i, i=1,\dots,n$ is spanned by the Mandelstam variables:
\begin{equation}
s_{ij}=(p_i+p_j)^2,\ \ \text{for}\ 1\leq i<j\leq n.
\end{equation}  
We introduce generalized Mandelstam variables as:
\begin{equation}
s_{I}=\left(\sum_{i\in I}p_i\right)^2=\sum_{\substack{i,j\in I\\i<j}}s_{ij}.
\end{equation}  
A convenient basis for this kinematic space is so-called ``planar variables"  
\begin{equation}
X_{i,j}=(p_i+p_{i+1}+\dots+p_j)=s_{i,i+1,\dots,j-1}\ \ \text{for}\ 1\leq i<j\leq n.
\end{equation}  
By definition, we can see that $X_{i,i+1}=X_{1,n}=0$. These variables are identified with the diagonals of an $n$-gon. \par
The relation of these variables and Mandelstam variables is 
\begin{equation}
s_{ij}=X_{i,j+1}+X_{i+1,j}-X_{i,j}-X_{i+1,j+1}.
\end{equation} 
Each planar $n$-point tree $\phi^p$ graph has $D_p$ propagators and there is a one-to-one correspondence between this graph and a complete $p$-angulation of the $n$-gon with $D_p$ diagonals.\par
Next, we introduce a planar scattering form for $\phi^p$ interactions. In the case of $\phi^3$, the scattering form is determined uniquely by demanding the projectivity of the form \cite{Arkani-Hamed:2017mur}. However, there is no notion of projectivity for this general interactions, then we need to choose a subset of graphs to determine a scattering form uniquely by its projectivity. This subset of graphs is obtained from the $Q$-compatible set of the complete $p$-angulations. Then we can define a planar scattering form for this subset \cite{Raman:2019utu}.  \par
%This subset is given as follows. First, we consider a reference graph $g$ and a corresponding $p$-angulation $P$. We operate $Q$-flips for this $P$ and make a sequence of $p$-angulations. These $p$-angulations make a subset and we call it as $Q$-compatible set "$Q(P)$". These $p$-angulations of the $Q(P)$ are the vertices of the accordiohedron. 
Let $g(Q(P))$ denote tree graphs for $\phi^p$ interactions in the $Q$-compatible set $Q(P)$ with propagators $X_{i_a,j_a}$ for $a=1,\cdots,D_p$. Then a $Q(P)$ dependent planar scattering form $\Omega_n^{P}$ is defined as
 \begin{equation}
 \label{eq:scatteringform}
\Omega_n^{P}=\sum_{g(Q(P))}\text{sign}(g(Q(P)))\bigwedge_{a=1}^{D_p}d\log{X_{i_a,j_a}}
\end{equation} 
where sign$g(Q(P))=\pm1$ and we sum over a $d$log form for every planar graphs in the subset $P$. The sign is determined as
 \begin{equation}
\text{sign}g(Q(P))=-\text{sign}g'(Q(P))
\end{equation} 
where $g'$ is a graph which obtained by the $Q$-flip of $g$. This rule make the form projective. \par
The important point is that any $Q$-compatible set of graphs does not exhaust all the graphs (or $p$-angulations). Then we need to sum all of these subsets with appropriate weight to obtain the scattering amplitude. To do this, we introduce ``a subset of primitive $p$-angulations" $\{P_1,\dots,P_I\}$:
\begin{itemize}
\item no two $p$-angulation of this subset are related to each other by cyclic permutations $\sigma$
\item all the other $p$-angulations are obtained by a cyclic permutations of $p$-angulation of this subset.
\end{itemize}
Then the $\phi^p$ tree amplitude is given as
\begin{equation}
\label{eq:amplitudesum}
M_{p,n}=\sum_{\sigma}\sum_{\text{primitive}\ P}\alpha_{p,n}^{P} m_{p,n}^{P,\sigma}
\end{equation} 
where $\alpha_{p,n}^{P}$ is a weight and $m_{p,n}^{P,\sigma}$ is the ratio part of the scattering form:
\begin{equation}
m_{p,n}^{P}=\left(\sum_{g(Q(P))}\frac{1}{\prod_{a=1}^{D_p}X_{i_a,j_a}}\right)
\end{equation} 
We call this $m_{p,n}^{P}$ as ``$Q(P)$-compatible amplitude".\par
Since the cyclic permutation of the primitive $p$-augulation does not change the relative configuration of diagonals, the weights depend only on primitive $p$-angulations. 
In the next section, we see how to embed the accordiohedron into the kinematic space and the relation between the canonical form of this kinematic accordiohedron and the scattering form. 
%-----------------------------------------------------------------------------------------------------------
\subsection{The kinematic accordiohedron}
The kinematic accordiohedron $\mathcal{AC}^{P}_{p,D_p}$ is a $D_p$-dim polytope defined as the intersection of the positive region:
\begin{equation}
\Delta_n=\{X_{ij}\geq0\  \text{for all}\ 1\leq i<j\leq n\}
\end{equation} 
with a hyperplane defined as follows:
\begin{equation}
\begin{split}
\label{eq:defofzccordiohedron}
H_n=\{&C_{ij}=-X_{i,j+1}-X_{i+1,j}+X_{i,j}+X_{i+1,j+1}\  \text{for}\ 1\leq i<j\leq n-1\\
&d_{r_i,s_i}=X_{r_i,s_i}\  \text{s.t.}\ (r_i,s_i)\ \text{is a complete triangulation}\}
\end{split}
\end{equation} 
where $C_{i,j}$ and $d_{r_i,s_i}$ are positive constants. The dimension of this hyperplane is given as 
\begin{equation}
d(H_n)=\frac{(n-2)(np-2n-p)}{2(p-2)}.
\end{equation} 
It is easy to check that these dimensions are correct. First, the dimension of $\Delta_n$ is $n(n-3)/2$.  Next, we consider the dimension of the hyperplane $H_n$. The number of the first line condition in \eqref{eq:defofzccordiohedron} is $(n-3)(n-2)/2$ as same as the associahedron case. When we $p$-angulate the $n$-point polygon, this polygon decomposes into $D_p+1$ $p$-gons. For each $p$-gon, we need to consider the complete triangulation. The number of triangulation of the $p$-gon is $p-2$ and there are $p-3$ diagonals. From this, the number of the second line conditions are 
\begin{equation}
(D_p+1)\times (p-3)=\frac{(n-2)(p-3)}{p-2}.
\end{equation} 
Then the total number of the constraints of \eqref{eq:defofzccordiohedron} is
\begin{equation}
\frac{(n-2)(p-3)}{p-2}+\frac{(n-3)(n-2)}{2}=\frac{(n-2)(np-2n-p)}{2(p-2)}.
\end{equation} 
This is the dimension of the hyperplane $H_n$. Finally, the dimension of the intersection $\mathcal{AC}^{P}_{p,D_p}=\Delta_n \cap H_n$ is
\begin{equation}
\frac{n(n-3)}{2}-\frac{(n-2)(np-2n-p)}{2(p-2)}=\frac{n-p}{p-2}=D_p.
\end{equation} 
\par
Since the accordiohedron $\mathcal{AC}^{P}_{p,D_p}$ is a simple polytope, the canonical form is a sum over its vertices \cite{Arkani-Hamed:2017tmz}. For each vertex $Z$, we denote its adjacent facets as $X_{i_a,j_a}=0$ for $a=1,\dots,D_p$. Then the canonical form of $\mathcal{AC}^{P}_{p,D_p}$ is written as
\begin{equation}
\label{eq:canonicalformdef}
\Omega(\mathcal{AC}^{P}_{p,D_p})=\sum_{\text{vertex}\ Z}\text{sign}(Z)\bigwedge_{a=1}^{D_p}d\log{X_{i_a,j_a}}
\end{equation} 
where sign$(Z)$ is evaluated on the ordering of the facets. 
This canonical form \eqref{eq:canonicalformdef} is equivalently the pullback of the scattering form \eqref{eq:scatteringform} to the subspace $H_n$. Then the ratio part of the canonical form (canonical function) corresponds to the $Q(P)$-compatible amplitude. From \eqref{eq:amplitudesum}, the tree amplitude of $\phi^p$ theory is given as the weighted sum of these $Q(P)$-compatible amplitudes. %This means that the canonical form of the kinematic accordiohedron determines the tree amplitude of $\phi^p$ theory. 
In the next section, we will see some explicit calculations to obtain the amplitudes from the kinematic accordiohedron.\par

%-----------------------------------------------------------------------------------------------------------
\subsection{Amplitudes from the accordiohedron}
\label{sec:Amplitudes from the accordiohedron}
\begin{figure}[t]
\begin{center}
\includegraphics[clip,width=17.0cm]{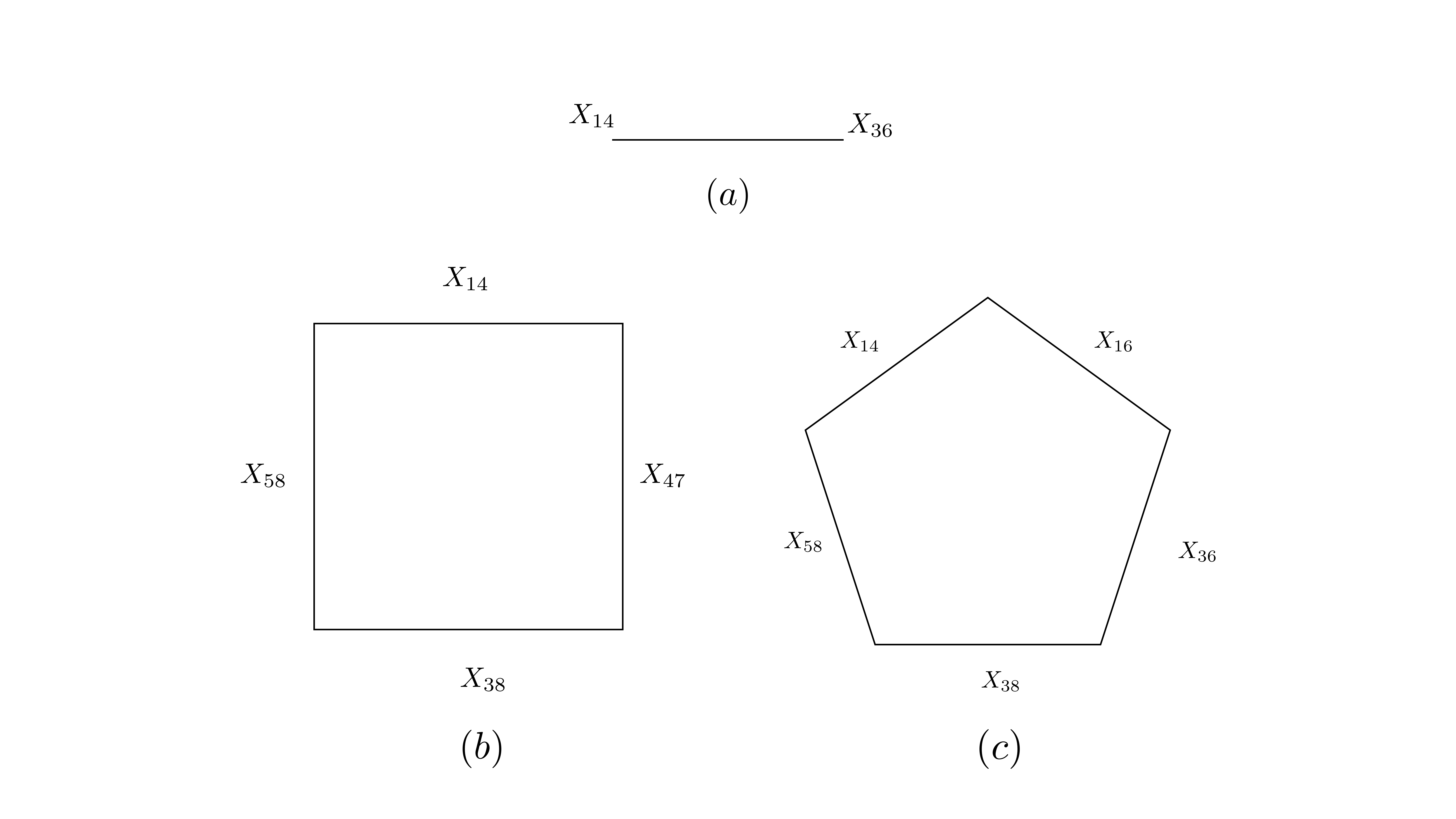}
 \caption{Stokes polytopes.}
 \label{fig:p4stokespolytope}
\end{center}
  \end{figure}

First, we see the simplest example: $p=4, n=6$ case. In this case, there is only one primitive quadrangulation, and all the others are obtained from the cyclic permutations. We choose $P=14$ as the reference quadrangulation, here $ij$ denotes the diagonal. The $Q$-compatible set of this reference quadrangulation is $Q(14)=\{(14,+),(36,-)\}$. The signs are determined that if a quadrangulation $P'$ is related to the reference quadrangulation $P$ by an odd (even) number of $Q$-flips, we associate $-(+)$ sign. Then the accordiohedron (stokes polytope) is a one-dimensional line as $(a)$ in Figure \ref{fig:p4stokespolytope}. Its canonical form is given by
\begin{equation}
\Omega_{p=4,n=6}^{14}=(d\ln{X_{14}}-d\ln{X_{36}}).
\end{equation} 
This stokes polytope locates inside the kinematic space with the following constraints
\begin{equation}
\begin{split}
s_{ij}&=-C_{ij},\ \text{for}\ 1\leq i<j\leq5\ \text{with} |i-j| \geq2\\
X_{13}&=d_{13}, X_{15}=d_{15}, \text{with}\ d_{13},d_{15}>0.
\end{split}
\end{equation}  
From these constraints, we can see that the planar kinematic variables satisfy
\begin{equation}
\begin{split}
X_{36}&=-X_{14}+\sum_{\substack{1\leq a <3\\4\leq b <6}}C_{ab}\geq0\\
X_{25}&=d_{15}-d_{13}+C_{14}+C_{13}\geq0.
\end{split}
\end{equation}  
Now we can pull back the canonical form as
\begin{equation}
\Omega_{4,6}^{14}=\left(\frac{1}{X_{14}}+\frac{1}{X_{36}}\right)dX_{14}= m_{4,6}^{14}dX_{14}.
\end{equation} 
It is easy to calculate other primitive quadrangulations cases $Q(25),Q(36)$. From \eqref{eq:amplitudesum}, the six-point amplitude is given as
\begin{equation}
M_{p=4,n=6}=\alpha_{p=4,n=6}^{14} (m_{4,6}^{14}+m_{4,6}^{25}+m_{4,6}^{36})=2\alpha_{4,6}^{14} \left(\frac{1}{X_{14}}+\frac{1}{X_{36}}+\frac{1}{X_{25}}\right),
\end{equation} 
where $\alpha_{4,6}^{14}$ is a weight and if we choose $\alpha_{4,6}^{14}=1/2$, this corresponds to the six-point amplitude. \par
Let us consider the $n=8$ case. In this case, there are two primitive quadrangulations $\{P_1,P_2\}$ and we take them to be $\{P_1=(14,58), P_2=(14,16)\}$. The $Q$-compatible set $Q(P_1)$ is given by
\begin{equation}
\label{eq:Qcompatible8pt}
Q(P_1)=\{(14,58,+),(14,47,-),(83,58,-),(83,47,+)\}.
\end{equation}  
Then the stokes polytope $\mathcal{S}^{P_1}$ is a two dimensional square as $(b)$ in Figure \ref{fig:p4stokespolytope}.
Similarly, $Q(P_2)$ is given by
\begin{equation}
\label{eq:Q2compatible8pt}
Q(P_2)=\{(14,16,+),(14,58,-),(36,16,-),(36,83,+),(58,83)\}.
\end{equation}  
We can see that the stokes polytope $\mathcal{S}^{P_2}$ is a two-dimensional pentagon  as $(c)$ in Figure \ref{fig:p4stokespolytope}. The constraints locating these stokes polytopes inside the kinematic space are given by
\begin{equation}
\begin{split}
\label{eq:p4n8constraints}
\mathcal{S}^{P_1}:\ \ \ \ \ \ s_{ij}&=-C_{ij},\ \text{for}\ 1\leq i<j\leq7\ \text{with} |i-j| \geq2\\
X_{13}&=d_{13}, X_{48}=d_{48}, X_{57}=d_{57}\\
\mathcal{S}^{P_2}:\ \ \ \ \ \ s_{ij}&=-C_{ij},\ \text{for}\ 1\leq i<j\leq7\ \text{with} |i-j| \geq2\\
X_{13}&=d_{13}, X_{46}=d_{46}, X_{68}=d_{68}.
\end{split}
\end{equation}  
From these constraints, the canonical forms become as
\begin{equation}
\begin{split}
\label{eq:p4n8canonicalform}
\Omega_{p=4,n=8}^{P_1}&=\left(\frac{1}{X_{14}X_{58}}+\frac{1}{X_{38}X_{47}}+\frac{1}{X_{14}X_{47}}+\frac{1}{X_{38}X_{58}}\right)d\ln{X_{14}}\wedge d\ln{X_{58}}\\
&=m_{4,8}^{P_1}d\ln{X_{14}}\wedge d\ln{X_{58}}\\
\Omega_{4,8}^{P_2}&=\left(\frac{1}{X_{14}X_{16}}+\frac{1}{X_{14}X_{58}}+\frac{1}{X_{36}X_{16}}+\frac{1}{X_{36}X_{83}}+\frac{1}{X_{58}X_{83}}\right)d\ln{X_{14}}\wedge d\ln{X_{16}}\\
&=m_{4,8}^{P_2}d\ln{X_{14}}\wedge d\ln{X_{16}}.
\end{split}
\end{equation}  
From \eqref{eq:amplitudesum}, the 8-point amplitude is given as
\begin{equation}
M_{p=4,n=8}=\alpha_{4,8}^{P_1}\left(m_{4,8}^{P_1}+\text{cyclic}\right)+\alpha_{4,8}^{P_2}\left(m_{4,8}^{P_2}+\text{cyclic}\right).
\end{equation} 
If we choose these weights as $\alpha_{4,8}^{P_1}=\frac{1}{3}, \alpha_{4,8}^{P_2}=\frac{1}{6}$, this corresponds to the 8-point amplitude. \par
\begin{figure}[t]
\begin{center}
\includegraphics[clip,width=17.0cm]{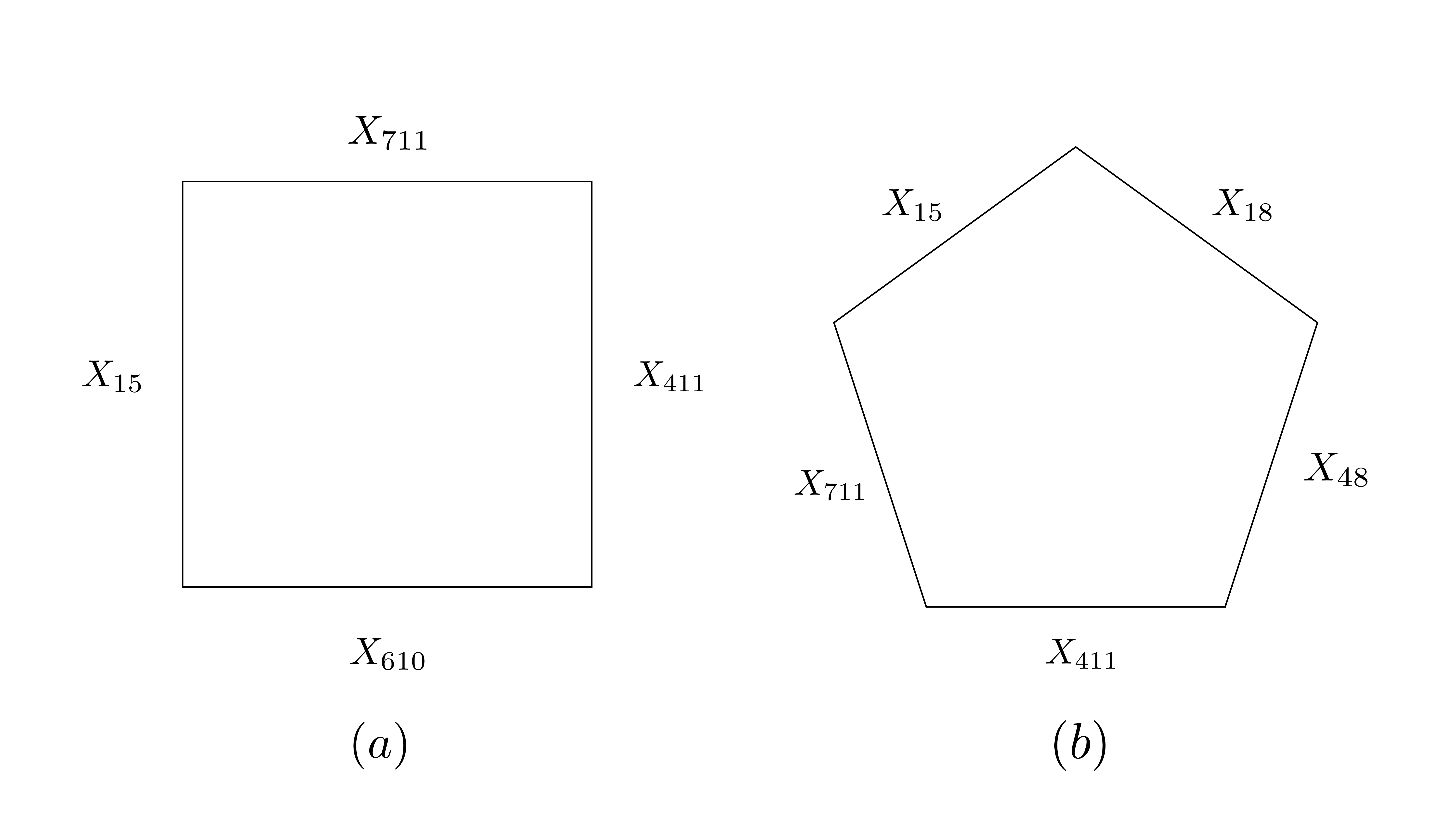}
 \caption{$p=5, n=11$ accordiohedra.}
 \label{fig:p5accordiohedron2}
\end{center}
  \end{figure}

Similarly we can calculate the amplitudes from the accordiohedron for general $p$ case. For example, we consider $p=5, n=11$ case. There are two $Q$-compatible sets $Q_5(P_1),Q_5(P_2)$:
\begin{equation}
\begin{split}
\label{eq:p5n11Qsets}
Q_5(P_1)&=\{(15,711),(411,711),(15,610),(411,610)\}\\
Q_5(P_2)&=\{(15,18),(18,48),(15,711),(411,711),(411,48)\}.
\end{split}
\end{equation}
The constraints are given as
\begin{equation}
\begin{split}
\label{eq:p5n11constraints}
Q_5(P_1):\ \ \ \ \ \ s_{ij}&=-C_{ij},\ \text{for}\ 1\leq i<j\leq10\  \text{with}\ |i-j| \geq 2\\
X_{13}&=d_{13},X_{35}=d_{35},X_{17}=d_{17},X_{57}=d_{57},X_{810}=d_{810},X_{710}=d_{710}.\\
Q_5(P_2):\ \ \ \ \ \ s_{ij}&=-C_{ij},\ \text{for}\ 1\leq i<j\leq10\  \text{with}\ |i-j| \geq 2\\
X_{13}&=d_{13},X_{35}=d_{35},X_{16}=d_{16},X_{68}=d_{68},X_{810}=d_{810},X_{110}=d_{110}.
\end{split}
\end{equation}
The shape of the accordiohedra for each $Q$-compatible set are given in Figure \ref{fig:p5accordiohedron2}.
The canonical forms for each accordiohedron are
\begin{equation}
\begin{split}
\label{eq:p5n11canonicalform}
\Omega_{p=5,n=11}^{P_1}&=\left(\frac{1}{X_{15}X_{610}}+\frac{1}{X_{411}X_{610}}+\frac{1}{X_{15}X_{711}}+\frac{1}{X_{411}X_{711}}\right)d\ln{X_{15}}\wedge d\ln{X_{610}}\\
&=m_{5,11}^{P_1}d\ln{X_{15}}\wedge d\ln{X_{610}}\\
\Omega_{5,11}^{P_2}&=\left(\frac{1}{X_{15}X_{18}}+\frac{1}{X_{18}X_{48}}+\frac{1}{X_{15}X_{711}}+\frac{1}{X_{411}X_{711}}+\frac{1}{X_{48}X_{411}}\right)d\ln{X_{15}}\wedge d\ln{X_{18}}\\
&=m_{5,11}^{P_2}d\ln{X_{15}}\wedge d\ln{X_{18}}.
\end{split}
\end{equation}  
From \eqref{eq:amplitudesum}, the 11-point amplitude is given as
\begin{equation}
M_{p=5,n=11}=\alpha_{5,11}^{P_1}\left(m_{5,11}^{P_1}+\text{cyclic}\right)+\alpha_{5,11}^{P_2}\left(m_{5,11}^{P_2}+\text{cyclic}\right).
\end{equation} 
If we choose these weights as $\alpha_{5,11}^{P_1}=\frac{3}{11}, \alpha_{5,11}^{P_2}=\frac{2}{11}$, this corresponds to the 11-point amplitude. \par
The important point is that to obtain the amplitudes, we need to choose the appropriate weights. In the next section, we see that these weights are determined by the factorization property of the accordiohedron.
%-----------------------------------------------------------------------------------------------------------
\section{Weights from the factorization}
In this section, we see that the weights can be obtained from the factorization property of the accordiohedron. In the planar $\mathcal{N}=4$ SYM, it is known that the geometric factorization of the amplituhedron implies physical factorization of scattering amplitudes \cite{Arkani-Hamed:2013jha}. This fact also holds for associahedron \cite{Arkani-Hamed:2017mur}, stokes polytope \cite{Banerjee:2018tun}, and accordiohedron \cite{Raman:2019utu}. First, we briefly review this factorization property of the accordiohedron. Then we see some explicit calculations of the weights.
\label{sec:weightfactorization}
\subsection{Factorization}
For any diagonal $(ij)$, we consider all primitive $p$-angulation $P$ which contains $(ij)$ and the kinematic accordiohedron $\mathcal{AC}_{p,n}^{P}$. The facet $X_{ij}=0$ of this accordiohedron is given as a product of lower dimensional accordiohedra
\begin{equation}
\label{eq:factorizationaccordiohedron}
\mathcal{AC}_{p,n}^{P}|_{X_{ij}=0}=\mathcal{AC}_{p,m}^{P_1}\times\mathcal{AC}_{p,n+2-m}^{P_2},
\end{equation} 
where $P_1$ is the $p$-angulation of the polygon $\{i,i+1,\dots,j\}$ and $P_2$ is the $p$-angulation of $\{j,j+1,\dots,n,1,\dots,i\}$. These $p$-angulation satisfy $P_1\cup P_2\cup (ij)=P$.\par
Next, we see that this geometric factorization implies the physical factorization of scattering amplitudes. The canonical form of the positive geometry satisfies the following property \cite{Arkani-Hamed:2017tmz}:
\begin{enumerate}
\item For any hyper-surface $H$ containing a boundary $\mathcal{B}$ of $\mathcal{A}$, the residue along $H$ is given by
\begin{equation}
\text{Res}_H\Omega(\mathcal{A})=\Omega(\mathcal{B}).
\end{equation} 
\item For any pair of positive geometries $\mathcal{A}$ and $\mathcal{B}$, we have
\begin{equation}
\Omega(\mathcal{A}\times \mathcal{B})=\Omega(\mathcal{A})\wedge\Omega(\mathcal{B}).
\end{equation} 
\end{enumerate}
By using these facts, we can obtain the relation 
\begin{equation}
\label{eq:factorization}
\text{Res}_{X_{ij}=0}\Omega(\mathcal{AC}_{p,n}^{P})=\Omega(\mathcal{AC}_{p,m}^{P_1})\wedge \Omega(\mathcal{AC}_{p,n+2-m}^{P_2}).
\end{equation} 
This factorization property implies the physical factorization of amplitudes. The diagonal $(ij)$ divides the $n$-gon into two polygons $\{i,i+1,\dots,j\}$ and $\{j,j+1,\dots,n,1,\dots,i\}$. By considering the kinematic accordiohedron associated to these polygons, we can obtain two sub-amplitudes $\{M_{|j-i+1|}, M_{n+2-(|j-i+1|)}\}=\{M_L,M_R\}$. Then \eqref{eq:factorization} implies that
\begin{equation}
\label{eq:factorization1}
M_n|_{X_{ij}=0}=M_L\frac{1}{X_{ij}}M_R.
\end{equation} 
 This is the physical factorization of the amplitudes. 
%-----------------------------------------------------------------------------------------------------------
\subsection{Determination of the weights}
In this section, we see that the weights can be determined from the factorization property of the accordiohedron. From \eqref{eq:factorizationaccordiohedron} and \eqref{eq:factorization}, the weights are constrained as
\begin{equation}
\sum_{P\in (ij)}\alpha_P=\sum_{P_L,P_R}\alpha_{P_L}\alpha_{P_R}.
\end{equation} 
The left hand side involves sum over all accordiohedra $\mathcal{AC}_{p,n}^{P}$ for which $P\in (ij)$ and the right hand side involves sum over $P_L$ and $P_R$ which range over all the $p$-angulations of the two polygons $\{i,i+1,\dots,j\}$ and $\{j,j+1,\dots,n,1,\dots,i\}$ respectively. \par
In addition to this, we can further constraint the weights from the factorization. By choosing the diagonal $(kl)\in\{i,i+1,\dots,j\}$, the polygon $\{i,i+1,\dots,j\}$ is divided into two polygons $\{k,k+1,\dots,l\}$ and $\{l,l+1,\dots,i,j,\dots,k\}$. By repeating the same procedure of the before section, we obtain 
\begin{equation}
\label{eq:factorization2}
M_{i,i+1,\dots,j}|_{X_{kl}=0}=M_{L_2}\frac{1}{X_{kl}}M_{R_2}.
\end{equation} 
where $M_{L_2}$ and $M_{R_2}$ are sub-amplitudes $\{M_{L_2},M_{R_2}\}=\{M_{|l-k+1|}, M_{i+j+1-|l-k+1|}\}$. From \eqref{eq:factorization1} and \eqref{eq:factorization2}, we obtain
\begin{equation}
M_n|_{X_{ij}=0,X_{kl}=0}=M_{L2}\frac{1}{X_{kl}}M_{R_2}\frac{1}{X_{ij}}M_R.
\end{equation} 
Then the weights satisfy 
 \begin{equation}
\sum_{P\in (ij),(kl)}\alpha_P=\sum_{P_{L_2},P_{R_2},P_R}\alpha_{P_{L_2}}\alpha_{P_{R_2}}\alpha_{P_R}.
\end{equation} 
The left hand side involves sum over all accordiohedra $\mathcal{AC}_{p,n}^{P}$ for which $P\in (ij),(kl)$ and the right hand side involves sum over $P_{L_2}, P_{R_2}$ and $P_R$ which range over all the $p$-angulations of the three polygons $\{k,k+1,\dots,l\}$, $\{l,l+1,\dots,i,j,\dots,k\}$ and $\{j,j+1,\dots,n,1,\dots,i\}$ respectively. Continuing this procedure, we can obtain 
\begin{equation}
M_n|_{X_{i_1j_1},X_{i_2j_2},\dots,X_{i_k,j_k}=0}=M_{L_k}\frac{1}{X_{i_kj_k}}M_{R_k}\frac{1}{X_{i_{k-1}j_{k-1}}}M_{R_{k-1}}\cdots M_{R_2}\frac{1}{X_{i_1j_1}}M_{R_1},
\end{equation} 
and
\begin{equation}
\label{eq:weightconstraints}
\sum_{P\in (i_1j_1),(i_2j_2),\dots,(i_kj_k)}\alpha_P=\sum_{P_{L_k},P_{R_k},\dots P_{R_1}}\alpha_{P_{L_k}}\alpha_{P_{R_k}}\cdots\alpha_{P_{R_1}}.
\end{equation} 
If the diagonals $(i_1j_1),(i_2j_2),\dots,(i_kj_k)$ make the complete $p$-angulation of $P$, the all weights of the right hand side are $1$. In this case, the constraints become simple as
\begin{equation}
\label{eq:weightconstraints2}
\sum_{P\in (i_1j_1),(i_2j_2),\dots,(i_kj_k)}\alpha_P=1.
\end{equation}
Since the complete $p$-angulation $(i_1j_1),(i_2j_2),\dots,(i_kj_k)$ is the vertex of the accordiohedron, the canonical form of the accordiohedron $\mathcal{AC}^P_{p,n}$ for $P\in (i_1j_1),\dots,(i_kj_k)$ has the term
 $\prod_{a=1}^k1/X_{i_a,j_a}$. Then the sum of the left-hand side of \eqref{eq:weightconstraints2} means that sum over all accordiohedra for which has the term $\prod_{a=1}^k1/X_{i_a,j_a}$ in its canonical form. From this, we can interpret the equation \eqref{eq:weightconstraints2} as the constraint that in the weighted sum of all canonical forms, each $\prod_{a=1}^k1/X_{i_a,j_a}$ appears exactly once. This is corresponding to the constraint of the weights obtained in \cite{Banerjee:2018tun, Raman:2019utu} by demanding that the weighted sum of canonical forms gives the full amplitude. In \cite{Raman:2019utu}, it was also proven that to satisfy this constraint, it is sufficient to impose these constraints for the primitive $p$-angulations. 
\begin{equation}
\sum_{i=1}^ln^i_P\alpha_P^i=1\ \ \ \text{for each primitive}\ 1\leq i \leq l
\end{equation}
where $i$ labels each primitive $p$-angulation, $n^i_P$ is number of times primitive $i$ appears in the vertices of all accordiohedra and $\alpha_P^i$ are the corresponding weights. Here we assumed that there are $l$ primitive $p$-angulations. The important point is that the derivation of this condition in \cite{Banerjee:2018tun, Raman:2019utu} relied on the form of scattering amplitudes. However, here we derive this formula from the factorization property of the accordiohedron.\par  
By using these constraints, we can determine the weights uniquely. Let us consider $p=4$ case. In the most simple $n=4$ case, the weight is trivial $\alpha_{4,4}=1$. Next we consider $n=6$ case. There is only way to divide the hexagon $\{1,2,\dots,6\}$ into two squares and both these have trivial weights, then \eqref{eq:weightconstraints} gives
\begin{equation}
\begin{split}
2\alpha_{4,6}&=1\\
\alpha_{4,6}&=\frac{1}{2}.
\end{split}
\end{equation}
In the $n=8$ case, we choose the diagonals $\{(47),(38)\}$ and $\{(38),(36)\}$. Then  the equation \eqref{eq:weightconstraints} gives two constraints
\begin{equation}
\begin{split}
2\alpha_{4,8}^{P_1}+2\alpha_{4,8}^{P_2}&=1\\
\alpha_{4,8}^{P_1}+4\alpha_{4,8}^{P_2}&=1.
\end{split}
\end{equation}
From this, we obtain the correct weights $\alpha_{4,8}^{P_1}=\frac{1}{3}, \alpha_{4,8}^{P_2}=\frac{1}{6}$. If we choose another diagonals, the result does not change.  For example if we choose the diagonals $\{(38)\}$ and $\{(14),(58)\}$, \eqref{eq:weightconstraints} gives
\begin{equation}
\begin{split}
2\alpha_{4,8}^{P_1}+5\alpha_{4,8}^{P_2}&=\frac{3}{2}\\
2\alpha_{4,8}^{P_1}+2\alpha_{4,8}^{P_2}&=1.
\end{split}
\end{equation}
These give the same weights. Finally we consider $n=10$ case. We choose the diagonals $\{(14),(510)\}$, $\{(14),(69)\}$, $\{(14),(49)\}$, $\{(14),(58)\}$, $\{(14),(510),(69)\}$, $\{(14),(49),(69)\}$, \\$\{(14),(510),(58)\}$. Then the equation \eqref{eq:weightconstraints} gives
\begin{equation}
\begin{split}
2\alpha^{P_1}+2\alpha^{P_2}+\alpha^{P_3}+\alpha^{P_4}+2\alpha^{P_5}+3\alpha^{P_6}+3\alpha^{P_7}&=\frac{3}{2}\\
2\alpha^{P_1}+4\alpha^{P_2}+3\alpha^{P_3}+3\alpha^{P_4}+2\alpha^{P_5}+2\alpha^{P_6}+2\alpha^{P_7}&=\frac{3}{2}\\
\alpha^{P_1}+5\alpha^{P_2}+3\alpha^{P_3}+2\alpha^{P_4}+3\alpha^{P_5}+3\alpha^{P_6}+2\alpha^{P_7}&=\frac{3}{2}\\
\alpha^{P_1}+3\alpha^{P_2}+\alpha^{P_3}+\alpha^{P_4}+4\alpha^{P_5}+3\alpha^{P_6}+3\alpha^{P_7}&=\frac{3}{2}\\
2\alpha^{P_1}+\alpha^{P_3}+\alpha^{P_4}+2\alpha^{P_6}+2\alpha^{P_7}&=1\\
\alpha^{P_1}+4\alpha^{P_2}+3\alpha^{P_3}+2\alpha^{P_4}+2\alpha^{P_5}+2\alpha^{P_6}&=1\\
\alpha^{P_1}+2\alpha^{P_2}+\alpha^{P_4}+2\alpha^{P_5}+2\alpha^{P_6}+2\alpha^{P_7}&=1,
\end{split}
\end{equation}
where $\alpha^{P_i}=\alpha_{4,10}^{P_i}$. From these equations, we can obtain 
\begin{equation}
\begin{split}
\alpha^{P_1}&=\frac{5}{24}, \alpha^{P_2}=\frac{1}{24}, \alpha^{P_3}=\frac{1}{24}, \alpha^{P_4}=\frac{1}{24}, \alpha^{P_5}=\frac{1}{12},
\alpha^{P_6}=\frac{1}{8}, \alpha^{P_7}=\frac{1}{8}.
\end{split}
\end{equation}
These are the correct weights.\par
We can similarly calculate the weights of the general $p$ case from the factorization. Let us consider $p=5, n=11$ case, we choose the diagonals $\{(15)\}$ and $\{(15),(711)\}$, \eqref{eq:weightconstraints} gives
\begin{equation}
\begin{split}
4\alpha_{5,11}^{P_1}+5\alpha_{5,11}^{P_2}&=2\\
3\alpha_{5,11}^{P_1}+\alpha_{5,11}^{P_2}&=1.
\end{split}
\end{equation}
From this, we obtain the correct weights $\alpha_{5,11}^{P_1}=\frac{3}{11}, \alpha_{5,11}^{P_2}=\frac{2}{11}$. 

%----------------------------------------------------------------------------------
\section{``BCFW"-like recursion relation}
\label{sec:recursion relation}
\subsection{General setups}
In this section, we construct a one-parameter recursion relation for $\phi^p$ amplitudes. This is just the generalization of the BCFW-like recursion relation of $\phi^3$ case \cite{Arkani-Hamed:2017mur, He:2018svj} to the $\phi^p$ case. The important point is that in the general $\phi^p$ cases, the recursion relation can be defined only for each $Q$-compatible set of graphs. We choose the $Q$-compatible set $Q(P)$ which is obtained from a reference $p$-angulation $P$ and consider the $D_p$ basis variables $X_{A_i}$. Here we construct the recursion relation for this $Q$-compatible set of graphs explicitly from a one-parameter deformation of the basis variables:
\begin{equation}
X_{A_i}\rightarrow\hat{X}_{A_i}:=zX_{A_i},\ \ \text{for}\ i=1,2,\cdots, D_p,
\end{equation}
where we do not change the constant $C_{ij}$. Following the same logic as the case of $\phi^3$, we consider the contour integral:
\begin{equation}
m^{P}_{p,n}(X,C)=\oint_{|z-1|=\epsilon}\frac{z^{D_p}dz}{z-1}m^{P}_{p,n}(\hat{X},C).
\end{equation}
Then the amplitude is given by the residue at $z=1$. From Cauchy theorem, this integral becomes as
\begin{equation}
\label{eq:integral}
m^{P}_{p,n}(X,C)=-\left(\text{Res}_{z=\infty}+\sum_{\text{finite poles}}\text{Res}_{z=z^*}\right)\frac{z^{D_p}dz}{z-1}m^{P}_{p,n}(\hat{X},C).
\end{equation}
Because of a $z^{D_p}$, the function $\frac{z^{D_p}}{z-1}m_{p,n}^{P}(zX,C)$ doesn't have a pole at $z=0$. One can verify that this function has no pole at infinity. We will see this in the later of this section. \par
Then we consider the residues at each finite pole. We denote the set of all non-basis variables $X_{B_i}$ which depend on the basis variables $X_{A_i}$ as
\begin{equation}
X_{B_i}=C_{B_i}+X_i+\sum_{j=1}^{D_p}\lambda_{i,j}X_{A_j},
\end{equation}
where $\lambda_{i,j}$ are real numbers and $C_{B_i}$ is a linear combination of constants $C$ and $X_i$ is a linear combination of unshifted planar variables. We denote $z_{B_i}$ as the  solutions of $\hat{X}_{B_i}(z)=0$. 
%Here we write the planar variables $X_{B_i}$ as
%\begin{equation}
%X_{B_i}=C_{B_i}+\sum_{j=1}^{D_p}\lambda_{i,j}X_{A_i},
%\end{equation}
%where $\lambda_{i,j}$ are real numbers and $C_{B_i}$ is a linear combination of constants $C$ and unshifted planar variables. 
The amplitude factorize at each physical pole as
\begin{equation}
\lim_{\hat{X}_{B_i}\rightarrow0}\hat{X}_{B_i}m_{p,n}^{P}(zX,C)=m^{P_L}_{a,\cdots,b-1,I}(z_{B_i}X,C)\times m^{P_R}_{I,b,\cdots,a-1}(z_{B_i}X,C),
\end{equation}
where $P_L\cup P_R\cup B_i=P$. Then the residue becomes as 
\begin{equation}
\text{Res}_{\hat{X}_{B_i}}\frac{z^{D_p}}{z-1}m_{p,n}^{P}(zX,C)=\frac{z_{B_i}^{D_p}}{\sum_{j=1}^{D_p}\lambda_{i,j}X_{A_j}(z_{B_i}-1)}m^{P_L}_{a,\cdots,b-1,I}(z_{B_i}X,C)\times m^{P_R}_{I,b,\cdots,a-1}(z_{B_i}X,C).
\end{equation}
By using $\hat{X}_{B_i}(z)=0$, the denominator becomes as
\begin{equation}
\sum_{j=1}^{D_p}\lambda_{i,j}X_{A_j}(z_{B_i}-1)=-X_{B_i}.
\end{equation}
Then we can obtain the recursion relation formula
\begin{equation}
\label{eq:BCFWrecursion}
m_{p,n}^{P}(X,C)=\sum_{B_i}\frac{z_{B_i}^{D_p}}{X_{B_i}}m^{P_L}_{a,\cdots,b-1,I}(z_{B_i}X,C)\times m^{P_R}_{I,b,\cdots,a-1}(z_{B_i}X,C)
\end{equation}
where $B_i$ runs over all the shifted planar variables. This is the BCFW-like recursion relation for the $Q$-compatible set of the amplitudes. For each $Q$-compatible set, we can apply this recursion relation. To obtain the amplitude, we need to sum over all primitive $p$-angulation and all cyclic permutation with the weight as \eqref{eq:amplitudesum}.\par
Next, we prove that there is no pole at infinity. We consider the behavior of function $\frac{z^{D_p}}{z-1}m_{p,n}^{P}(zX,C)$ at $z\rightarrow \infty$. The canonical function $m_{p,n}^{P}(X,C)$ of the accordiohedron have the form
\begin{equation}
m_{p,n}^{P}(X,C)=\sum_{\text{vertices}}\frac{1}{\prod^{D_p} X}.
\end{equation}
Since the dimension of the accordiohedron is $D_p$, there are no planar variables that are linear independent with basis variables. Then the canonical function has only terms of $\mathcal{O}\left(\frac{1}{z^{D_p}}\right)$. Then at $z\rightarrow \infty$, 
\begin{equation}
\lim_{z\rightarrow \infty} m_{p,n}^{P}(X,C)\sim\frac{1}{z^{D_p}}m_{p,n}^{P}(X,0).
\end{equation}
When we fix $i$ and set $D_p\ C_{ij}=0$, the canonical function of the accordiohedron vanishes as,
\begin{equation}
\label{eq:softfact}
m^{P}_{p,n}(zX,0)\rightarrow 0.
\end{equation}
We can easily verify this from a property of the canonical form of the accordiohedron. After we fix $i$ and set $D_p\ C_{ij}=0$, all facets of the accordiohedron pass through the origin. It means that the accordiohedron shrinks to the origin and the canonical form vanish. Without using the geometric property of the accordiohedron, we can also prove this from the factorization. First, we can see that all $2p-2$-point amplitudes vanish at $C=0$,
\begin{equation}
m^{P=(1p)}_{p,2p-2}(zX,0)=\frac{1}{X_{1p}}+\frac{1}{X_{p-1,2p-2}}=\frac{1}{X_{1p}}+\frac{1}{0-X_{1p}}=0.
\end{equation}
Here we use the equation
\begin{equation}
X_{1p}+X_{p-1,2p-2}=\sum_{\substack{1\leq a <p-1\\ p \leq b <2p-2 }}C_{ab} \rightarrow 0.
\end{equation}
We can see this for the other case obtained from the cyclic permutation. When lower-point amplitudes vanish with $C=0$, any residue of the amplitude in the factorization vanishes. From these facts, we can verify that the canonical function of the accordiohedron vanishes at $C=0$. Then the function $\frac{z^{D_p}}{z-1}m_{p,n}^{P}(zX,C)$ has no pole at $z\rightarrow \infty$. \par
In \cite{Yang:2019esm}, this condition is proved for general ABHY polytopes. In this proof, they used the fact that the ABHY-polytopes are "$C$-independent" polytopes. Since the accordiohedra are not $C$-independent, we cannot apply that proof to this accordiohedra. 
\par
The absence of the pole at infinity is related to the interesting properties of amplitudes. For example, in Yang-Mills theory, this fact is explained from the ``dual conformal symmetry". In this $\phi^p$ case, the $Q$-compatible sets of amplitudes have this property. It is thus interesting to see an analog of this hidden symmetry for each $Q$-compatible set of amplitudes, not all amplitudes.\par 
Here we compare this BCFW-like recursion with the BCFW recursion relation \cite{Britto:2004ap, Britto:2005fq}. Under the BCFW shift of the external momenta, there are two contributions: residue parts of the finite pole and a boundary term that comes from the pole at $z\rightarrow \infty$. Since the boundary term does not vanish in general $\phi^p$ case, we need to consider not only the finite pole parts but also the boundary term. For example, the BCFW recursion relation with the boundary term for the $\phi^4$ case has constructed in \cite{Feng:2009ei}. The systematic way to determine this boundary term has constructed in \cite{Feng:2014pia}. Unlike this BCFW recursion, the BCFW-like recursion which has constructed from the positive geometry has no contribution from the boundary term. From this, the BCFW-like recursion is simpler than the usual BCFW recursion relation for the general $\phi^p$ case. 
We see some examples of this recursion relation for $p=4,5,6$ cases. 
%---------------------------------------------------------------------------------------------------------------------------
\subsection{Explicit calculations}
$\bold{p=4, n=6}$
\par We choose $P=(36)$ as the primitive quadrangulation and denote $X_{36}$ as the basis variable. Other variables are given as 
\begin{equation}
\begin{split}
X_{14}&=-X_{36}+C_{14}+C_{24}+C_{15}+C_{25},\\
X_{25}&=d_{15}+C_{14}-d_{13}+C_{13}.
\end{split}
\end{equation}
The term $m^{36_L}_{345I}\times m^{36_R}_{I612}$ is the residues at the pole $\hat{X}_{14}:=C_{14}+C_{24}+C_{15}+C_{25}-z_{14}X_{36}=0$. Here $z_{ab}$ is the solution of the equation $\hat{X}_{ab}(z)=0$. Then
\begin{equation}
z_{14}=\frac{C_{14}+C_{24}+C_{15}+C_{25}}{X_{36}}.
\end{equation}
Since four-point amplitudes are trivial $m^{36_{L,R}}_4=\pm1$, we can obtain the $Q(36)$-compatible amplitude:
\begin{equation}
m^{36}_{p=4,6}=\frac{z_{14}}{X_{14}}=\frac{(C_{14}+C_{24}+C_{15}+C_{25})}{X_{36}X_{14}}=\frac{1}{X_{14}}+\frac{1}{X_{36}}.
\end{equation}
If we choose $X_{14}$ as the basis, we can obtain the same result. Similarly, we can compute other primitive quadrangulations obtained from the cyclic permutations. The results are given as
\begin{equation}
\begin{split}
m^{25}_{4,6}&=\frac{1}{X_{25}}+\frac{1}{X_{14}}\\
m^{14}_{4,6}&=\frac{1}{X_{36}}+\frac{1}{X_{25}}.
\end{split}
\end{equation}
Sum of these terms with appropriate weight corresponds to the six-point amplitude,
\begin{equation}
M_{4,6}=\alpha_{4,6}^{14}(m^{36}_{4,6}+m^{25}_{4,6}+m^{14}_{4,6}).
\end{equation}
%---------------------------------------------------------------------------------------------------------------------------
$\bold{p=4, n=8}$  
\par Next, we obtain the eight-point amplitude from six-point amplitudes. We have seen in the section \ref{sec:Amplitudes from the accordiohedron} that there are two primitive $P_1,P_2$. First, we choose $P_1=(14,58)$ as the primitive quadrangulation, then the $Q$-compatible set is given as \eqref{eq:Qcompatible8pt}. If we choose $X_{14},X_{58}$ as the basis, we need to sum over residues at $\hat{X}_{38}=0,\hat{X}_{47}=0$.  They correspond to the factorizations:
\begin{equation}
\hat{X}_{38}=0: \ m^{{P_1}_L}_{34567I}\times m^{{P_1}_R}_{I812} ,\ \ \hat{X}_{47}=0:\ m^{{P_1}_L}_{456I}\times m^{{P_1}_R}_{I78123}.
\end{equation}
From the results of the four-point and six-point amplitude, we have:
\begin{equation}
\begin{split}
m^{{P_1}_R}_{I812}&=1,\ m^{{P_1}_L}_{34567I}=\frac{1}{\hat{X}_{58}(z_{38})}+\frac{1}{\hat{X}_{47}(z_{38})}\\
m^{{P_1}_L}_{456I}&=1,\ m^{{P_1}_R}_{I78123}=\frac{1}{\hat{X}_{14}(z_{47})}+\frac{1}{\hat{X}_{38}(z_{47})}.
\end{split}
\end{equation}
This gives the result for $Q(P_1)$-compatible amplitude:
\begin{equation}
\begin{split}
\label{eq:8ptrecursion}
m^{P_1}_{4,8}&=\frac{z_{38}^2}{X_{38}}\left(\frac{1}{\hat{X}_{58}(z_{38})}+\frac{1}{\hat{X}_{47}(z_{38})}\right)+\frac{z_{47}^2}{X_{47}}\left(\frac{1}{\hat{X}_{14}(z_{47})}+\frac{1}{\hat{X}_{38}(z_{47})}\right)\\
&=\frac{A_{38}^2}{X_{38}X_{14}}\left(\frac{1}{X_{58}A_{38}}+\frac{1}{(A_{47}X_{14}-A_{38}X_{58})}\right)+\frac{A_{47}^2}{X_{47}X_{58}}\left(\frac{1}{X_{14}A_{47}}+\frac{1}{(A_{38}X_{58}-A_{47}X_{14})}\right)
\end{split}
\end{equation}
where $A_{38}, A_{47}$ are given as 
\begin{equation}
\label{eq:a38a47}
A_{38}=\sum_{i=4}^7(C_{1i}+C_{2i}), A_{47}=d_{57}+d_{48}+C_{47}.
\end{equation} 
From the straightforward calculation, we can see that
\begin{equation}
m^{P_1}_{4,8}=\frac{1}{X_{14}X_{47}}+\frac{1}{X_{38}X_{47}}+\frac{1}{X_{14}X_{58}}+\frac{1}{X_{38}X_{58}}.
\end{equation}
Next, we consider the $P_2=(14,16)$ and the $Q(P_2)$-compatible set is given as \eqref{eq:Q2compatible8pt}. We choose $X_{14},X_{16}$ as the basis. We need to sum over residues at $\hat{X}_{58}=0,\hat{X}_{38}=0,\hat{X}_{36}=0$. Each residue is given as 
\begin{equation}
\begin{split}
\hat{X}_{58}&=0:\ \ m^{{P_2}_L}_{567I}\times m^{{P_2}_R}_{I81234}=\frac{A_{58}A_{38}}{X_{14}(A_{38}X_{16}-A_{58}X_{14})(A_{58}-X_{16})},\\
\hat{X}_{38}&=0:\ \ m^{{P_2}_L}_{34567I}\times m^{{P_2}_R}_{I812}=\frac{A_{36}A_{38}^2}{(X_{14}-A_{38})(A_{38}X_{16}-A_{58}X_{14})(A_{36}X_{14}-A_{38}X_{14}+A_{38}X_{16})},\\
\hat{X}_{36}&=0:\ \ m^{{P_2}_L}_{345I}\times m^{{P_2}_R}_{I67812}=\frac{(A_{36}-A_{58})(A_{36}+A_{38}-A_{58})}{X_{16}(A_{36}+A_{38}-A_{58}-X_{14}+X_{16})(A_{36}X_{14}-A_{38}X_{14}+A_{38}X_{16})}\\
\end{split}
\end{equation}
where $A_{58}, A_{36}$ are given as 
\begin{equation}
\label{eq:a58a36}
A_{58}=\sum_{\substack{1\leq a <5\\6\leq b <8}}C_{ab},\ A_{36}=\sum_{\substack{3\leq a <5\\6\leq b <8}}C_{ab}.
\end{equation}
We can see that 
\begin{equation}
m^{P_2}_{4,8}=\frac{1}{X_{14}X_{16}}+\frac{1}{X_{14}X_{58}}+\frac{1}{X_{36}X_{16}}+\frac{1}{X_{36}X_{83}}+\frac{1}{X_{58}X_{83}}.
\end{equation}
From these results, we can obtain the eight-point amplitude
\begin{equation}
M_{4,8}=\alpha^{P_1}_{4,8}m^{P_1}_{4,8}+\alpha^{P_2}_{4,8}m^{P_2}_{4,8}+\text{cyclic}
\end{equation}
where $\alpha^{P_1}_{4,8}=\frac{1}{3}$ and $\alpha^{P_2}_{4,8}=\frac{1}{6}$. These weights are obtained from the factorization.\\ \\
%------------------------------------------------------------
$\bold{p=4, n=10}$  \par
In this case, there are 7 primitive $\{P_1,P_2,\dots,P_7\}$. For each $P_i$, we need to apply the recursion relation to obtain the 10-point amplitude. First we consider $P_1=(14,510,69)$, the $Q$-compatible set is given as
\begin{equation}
\begin{split}
\label{eq:qsetn10}
Q(P_1)&=\{(14,510,69,+),(310,510,69,-),(14,49,69-),(14,510,58,-),\\
&(14,49,58,+),(310,510,58,+),(310,49,69,+),(310,49,58,-).
\end{split}
\end{equation}
Once we choose $X_{14},X_{510},X_{69}$ as the basis, we need to sum over residues at $\hat{X}_{310}=0,\hat{X}_{49}=0,\hat{X}_{58}=0$. 
%We can check that these residues correspond to the canonical functions which are obtained from the triangulation of the three-dimensional stokes polytope. 
Here we see the one example: the residue of $\hat{X}_{310}=0$. The shift parameter is given as
\begin{equation}
\hat{X}_{310}=z_{310}X_{14}+\sum_{\substack{1\leq a<3\\4\leq b<10}}C_{ab}=0\ \rightarrow \ z_{310}=\sum_{\substack{1\leq a<3\\4\leq b<10}}C_{ab}/X_{14}
\end{equation}
This residue corresponds to the factorization: 
\begin{equation}
m^{{P_1}_L}_{3456789I}\times m^{{P_1}_R}_{I1012}.
\end{equation}
By using the eight point result, 
\begin{equation}
\begin{split}
&\frac{z_{310}^3}{X_{310}}m^{P_1}_{3456789I}\times m^{P_1}_{I1012}\\
&=\frac{A_{310}^3}{X_{310}X_{14}^3}\left(\frac{1}{\hat{X}_{510}\hat{X}_{58}}+\frac{1}{\hat{X}_{49}\hat{X}_{58}}+\frac{1}{\hat{X}_{510}\hat{X}_{69}}+\frac{1}{\hat{X}_{49}\hat{X}_{69}}\right)\\
&=\frac{A_{310}A_{49}A_{58}X_{14}}{X_{310}X_{510}(A_{49}X_{14}-A_{310}X_{510})X_{69}(A_{58}X_{14}-A_{310}X_{69})}
\end{split}
\end{equation}
where
\begin{equation}
A_{310}=\sum_{\substack{1\leq a <3 \\ 4\leq b<10}}C_{ab},\ A_{58}=d_{68}+d_{59}+C_{59},\ A_{49}=d_{59}+d_{410}+C_{410}.
\end{equation}
Now it is easy to compute other residues.\\ \\
%----------------------------------------------------------------------------------------------------------------------- 
$\bold{p=5, n=8}$\\
In this case, there is one primitive $p$-angulation and we choose it as $P=(15)$. Once we choose $X_{15}$ as the basis, then $X_{48}$ is given as
 \begin{equation}
X_{48}=\sum_{i=1}^3(C_{i5}+C_{i6}+C_{i7}).
\end{equation}
The only term $m^{P_L}_{1234I}\times m^{P_R}_{I5678}$ is the residues at the pole $\hat{X}_{48}=0$. As same as the case of $\phi^4$, we can obtain the $Q(15)$-compatible amplitude:
\begin{equation}
\label{eq:n8p5Q1}
m^{15}_{p=5,n=8}=\frac{z_{48}}{X_{48}}=\frac{1}{X_{15}}+\frac{1}{X_{48}},
\end{equation}
where $z_{48}=\sum_{i=1}^3(C_{i5}+C_{i6}+C_{i7})/X_{15}$. The eight-point amplitude is given as the sum of all cyclic permutations of \eqref{eq:n8p5Q1}.\\ \\
%----------------------------------------------------------------------------------------------------------------------- 
$\bold{p=5, n=11}$\\
There are two primitive $p$-angulations $P_1=(15,711),P_2=(15,18)$ and two $Q$-compatible sets given as \eqref{eq:p5n11Qsets}.
First we consider $P_1$ and we choose $X_{15},X_{711}$ as the basis. We need to sum over residues at $\hat{X}_{411}=0, \hat{X}_{610}=0$. The residues at the pole $\hat{X}_{411}=0$ is $m^{{P_1}_L}_{45678910I}\times m^{{P_1}_R}_{I11123}$ and at the pole $\hat{X}_{610}=0$ is $m^{{P_1}_L}_{6789I}\times m^{{P_1}_R}_{I101112345}$. These residues are given as
\begin{equation}
\begin{split}
\hat{X}_{411}&=0:\ \ \frac{z_{411}^2}{X_{411}}\left(\frac{1}{\hat{X}_{117}(z_{411})}+\frac{1}{\hat{X}_{610}(z_{411})}\right)=\frac{A_{411}^2}{X_{411}X_{15}}\left(\frac{1}{A_{411}X_{117}}+\frac{1}{(A_{610}X_{15}-A_{411}X_{711})}\right)\\
\hat{X}_{610}&=0:\ \ \frac{z_{610}^2}{X_{610}}\left(\frac{1}{\hat{X}_{15}(z_{610})}+\frac{1}{\hat{X}_{411}(z_{610})}\right)=\frac{A_{610}^2}{X_{711}X_{610}}\left(\frac{1}{A_{610}X_{15}}+\frac{1}{(A_{411}X_{711}-A_{610}X_{15})}\right),
\end{split}
\end{equation}
where $A_{411},A_{610}$ are given as
\begin{equation}
\begin{split}
\label{eq:a411a610}
A_{411}=\sum_{\substack{1\leq a <4\\5\leq b<11}}C_{ab},\ A_{610}=\sum_{\substack{1\leq a <6\\7\leq b<11}}C_{ab}+C_{610}-d_{17}+d_{710}.
\end{split}
\end{equation}
The $Q(P_1)$-compatible amplitude $m^{P_1}_{p=5,n=11}$ is given as a sum of these two residues. From the straightforward calculation, we can see that 
\begin{equation}
m^{P_1}_{5,11}=\frac{1}{X_{15}X_{610}}+\frac{1}{X_{411}X_{610}}+\frac{1}{X_{15}X_{711}}+\frac{1}{X_{411}X_{711}}.
\end{equation}
Similarly we can compute the $P_2$ case. We choose $X_{15}, X_{18}$ as the basis. There are three residues at poles $\hat{X}_{48}=0,\hat{X}_{411}=0,\hat{X}_{711}=0$. These residues are given as
\begin{equation}
\begin{split}
\hat{X}_{48}=0:\ \ &\frac{z_{48}^2}{X_{48}}\left(\frac{1}{\hat{X}_{18}(z_{48})}+\frac{1}{\hat{X}_{411}(z_{48})}\right)=\frac{A_{48}(A_{48}-A_{411})}{X_{18}X_{48}(A_{48}X_{15}+A_{411}X_{18}-A_{411}X_{18})}\\
\hat{X}_{411}=0:\ \ &\frac{z_{411}^2}{X_{411}}\left(\frac{1}{\hat{X}_{48}(z_{411})}+\frac{1}{\hat{X}_{711}(z_{411})}\right)\\
&=\frac{A_{411}^2(A_{711}+A_{48}-A_{411})}{(A_{711}X_{15}-A_{411}X_{18})(A_{48}X_{15}+A_{411}X_{18}-A_{411}X_{15})X_{411}}\\
\hat{X}_{711}=0:\ \ &\frac{z_{711}^2}{X_{711}}\left(\frac{1}{\hat{X}_{15}(z_{711})}+\frac{1}{\hat{X}_{411}(z_{711})}\right)=\frac{A_{411}A_{711}}{X_{15}X_{711}(A_{411}X_{18}-A_{711}X_{15})},
\end{split}
\end{equation}
where $A_{48},A_{711}$ are given as
\begin{equation}
\begin{split}
\label{eq:a48a711}
A_{48}=\sum_{\substack{1\leq a <4\\5\leq b<8}}C_{ab},\ A_{711}=\sum_{\substack{1\leq a <7\\8\leq b<11}}C_{ab}.
\end{split}
\end{equation}
The $Q(P_2)$-compatible amplitude $m^{P_2}_{p=5,n=11}$ is
\begin{equation}
m^{P_2}_{5,11}=\frac{1}{X_{15}X_{18}}+\frac{1}{X_{18}X_{48}}+\frac{1}{X_{15}X_{711}}+\frac{1}{X_{411}X_{711}}+\frac{1}{X_{48}X_{411}}.
\end{equation}
Then the 10-point amplitude is given as a sum of these two compatible amplitudes with weights.
\\ \\
%----------------------------------------------------------------------------------------------------------------------- 
$\bold{p=6, n=14}$\\
There are three primitive $p$-angulations $P_1=(16,914),P_2=(16,110),P_3=(16,813)$ and three $Q$-compatible sets are given as
\begin{equation}
\begin{split}
\label{eq:qsetn14p6}
Q_6(P_1)&=\{(16,914),(514,914),(16,813),(514,813)\}\\
Q_6(P_2)&=\{(16,813),(514,813),(16,712),(514,712)\}\\
Q_6(P_3)&=\{(16,110),(16,914),(110,510),(510,514),(514,914)\}.
\end{split}
\end{equation}
Even in this $p=6$ case, we can compute each residue as same as lower $p$ case. Here we consider only $P_1$ case. There are two residues at $\hat{X}_{514}=0, \hat{X}_{813}=0$. These residues are given as 
\begin{equation}
\begin{split}
&\hat{X}_{514}=0:\ \ m^{{P_1}_L}_{5678910111213I}\times m^{{P_1}_R}_{I141234}=\left(\frac{1}{\hat{X}_{914}(z_{514})}+\frac{1}{\hat{X}_{813}(z_{514})}\right),\\
&\hat{X}_{813}=0:\ \ m^{{P_1}_L}_{89101112I}\times m^{{P_1}_R}_{I13141234567}=\left(\frac{1}{\hat{X}_{16}(z_{813})}+\frac{1}{\hat{X}_{514}(z_{813})}\right).\\
\end{split}
\end{equation}
Then the $Q_6(P_1)$-compatible amplitude is 
\begin{equation}
\begin{split}
m^{P_1}_{5,11}=&\frac{z_{514}^2}{X_{514}}\left(\frac{1}{\hat{X}_{914}(z_{514})}+\frac{1}{\hat{X}_{813}(z_{514})}\right)+\frac{z_{813}^2}{X_{813}}\left(\frac{1}{\hat{X}_{16}(z_{813})}+\frac{1}{\hat{X}_{514}(z_{813})}\right)\\
=&\frac{A_{514}^2}{X_{514}X_{16}}\left(\frac{1}{A_{514}X_{914}}+\frac{1}{A_{813}X_{16}-A_{514}X_{914}}\right)+\frac{A_{813}^2}{X_{813}X_{914}}\left(\frac{1}{A_{914}X_{16}}+\frac{1}{A_{514}X_{914}-A_{813}X_{16}}\right)\\
=&\frac{1}{X_{16}X_{914}}+\frac{1}{X_{514}X_{914}}+\frac{1}{X_{16}X_{813}}+\frac{1}{X_{514}X_{813}}
\end{split}
\end{equation}
where
\begin{equation}
A_{514}=\sum_{\substack{1\leq a<5\\6\leq b<14}}C_{ab},\ \ A_{813}=d_{814}+d_{913}+C_{813}.
\end{equation}
 There is no difficulty to compute other compatible amplitudes.

%-----------------------------------------------------------------------------------------------------------
\subsection{Recursion from the triangulation}
\label{sec:triangulation}
In this section, we interpret the results of the recursion relation as triangulations of accordiohedra. We see some examples for $p=4,5,6$ cases and obtain the general $n$-point formula.\\ \\
$\bold{p=4, n=8}$\par
\begin{figure}[t]
\begin{center}
\includegraphics[clip,width=17.0cm]{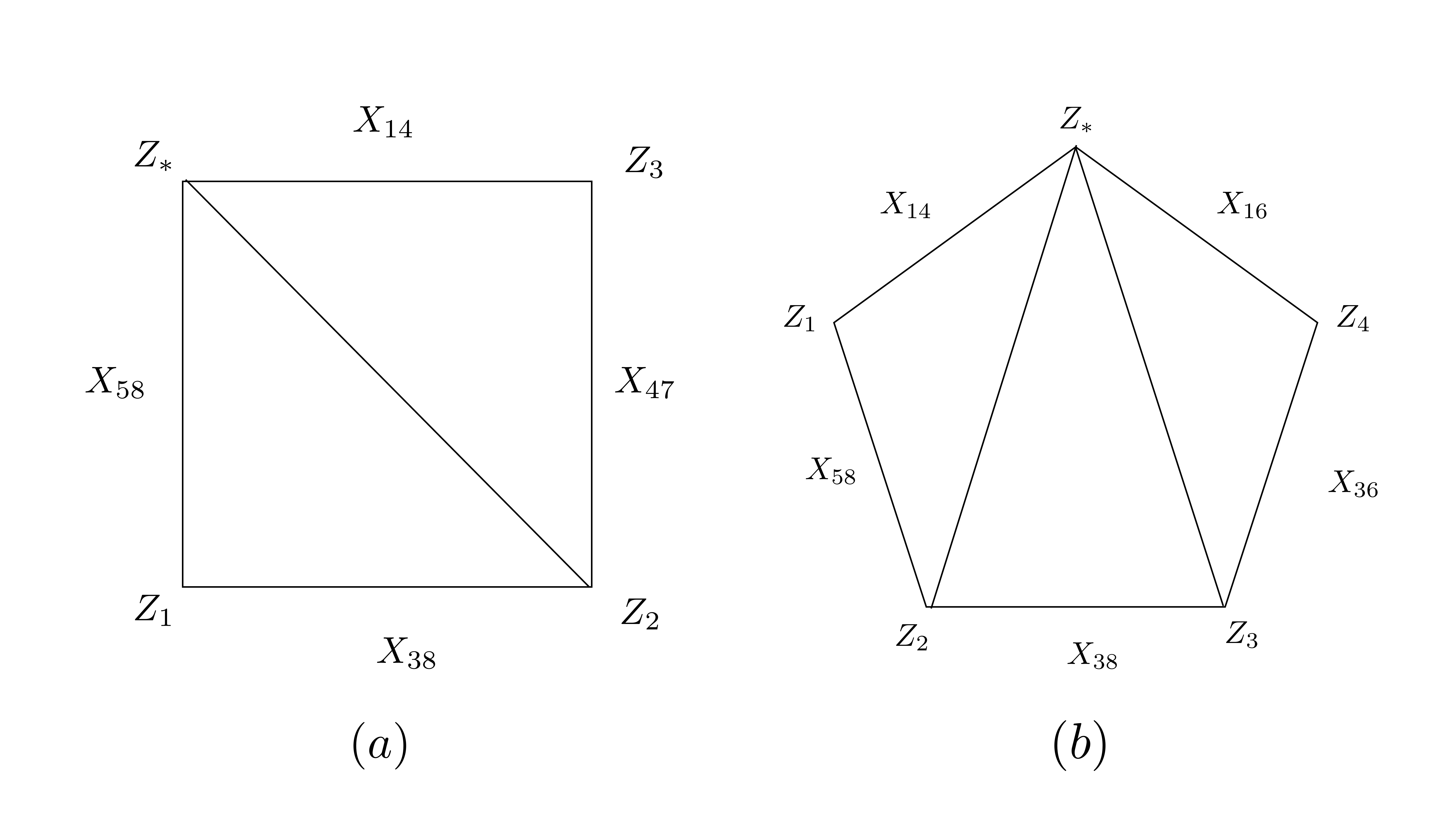}
 \caption{Triangulation of $p=4, n=8$ stokes polytopes.}
 \label{fig:stokespolytope}
\end{center}
  \end{figure}

We have seen in section \ref{sec:Amplitudes from the accordiohedron} that there are two stokes polytopes: one is a two dimensional square $\mathcal{S}^{P_1}$ and the other is a two-dimensional pentagon $\mathcal{S}^{P_2}$. First, we consider the triangulation of the $\mathcal{S}^{P_1}$. The constraints defining it inside the kinematic space are given in \eqref{eq:p4n8constraints}. To consider the triangulation, we use the coordinate for vertices of the stokes polytope. Once we choose a basis, the coordinates of any vertex can be obtained by solving $D_p$ linear equations of the form $X_{a,b}=0$. We choose the basis as $\{X_{14},X_{58}\}$ and let us denote the vertices 
\begin{equation}
\{X_{58},X_{14}\},\{X_{38},X_{58}\},\{X_{38},X_{47}\},\{X_{47},X_{14}\} \rightarrow \mathcal{Z}^{P_1}_*, \mathcal{Z}^{P_1}_1, \mathcal{Z}^{P_1}_2, \mathcal{Z}^{P_1}_3.
\end{equation}  
For example, a vertex $\mathcal{Z}^{P_1}_2$ can be obtained by solving
\begin{equation}
\begin{split}
-X_{14}+\sum_{\substack{1\leq a <3\\ 1\leq b <8}}C_{ab}=0,\ \ -X_{58}+d_{57}+d_{48}+C_{47}=0.
\end{split}
\end{equation}  
These equations give $\mathcal{Z}^{P_1}_2=(\sum_{\substack{1\leq a <3\\ 1\leq b <8}}C_{ab},d_{57}+d_{48}+C_{47})$. The direct computation gives
\begin{equation}
\begin{split}
\label{eq:coordinate8pt}
\mathcal{Z}^{P_1}_*&=(0,0), \ \mathcal{Z}^{P_1}_1=(\sum_{\substack{1\leq a <3\\ 1\leq b <8}}C_{ab},0), \\
\mathcal{Z}^{P_1}_2&=(\sum_{\substack{1\leq a <3\\ 1\leq b <8}}C_{ab},d_{57}+d_{48}+C_{47}),\  \mathcal{Z}^{P_1}_3=(0,d_{57}+d_{48}+C_{47}).
\end{split}
\end{equation}  
Then we introduce affine coordinate, $Z:=(1,\mathcal{Z})$ and $Y=(1,\bold{X})$, where $\bold{X}$ denotes the basis. We triangulate $\mathcal{S}^{P_1}$ into two triangles as $(a)$ in Figure \ref{fig:stokespolytope}.
  The canonical function of a simplex of dimension $(n-4)/2$ is given as
\begin{equation}
[Z_0,Z_1,\dots,Z_{\frac{n-4}{2}}]_{P}:=\frac{\langle Z_0Z_1\cdots Z_{\frac{n-4}{2}}\rangle_{P}^{\frac{n-4}{2}}}{\prod_{i=0}^{\frac{n-4}{2}}\langle Y Z_0Z_1\cdots \hat{Z}_i\cdots Z_{\frac{n-4}{2}}\rangle_{P}},
\end{equation}
where $\hat{Z}_i$ means the omission of $Z_i$. We use the notation 
\begin{equation}
\langle Z_0Z_1\cdots Z_{i}\rangle_{P}:=\epsilon_{IJ\cdots K}\left(Z^{P}_0\right)^I\left(Z^{P}_1\right)^J\cdots \left(Z^{P}_i\right)^K.
\end{equation}
From this formula, the canonical function of the $\mathcal{S}^{P_1}$ is given as
\begin{equation}
\begin{split}
&[Z_*,Z_1,Z_2]_{P_1}+[Z_*,Z_2,Z_3]_{P_1}\\
&=\frac{\langle Z_*Z_1Z_2\rangle
_{P_1}^2}{\langle YZ_*Z_1\rangle_{P_1}\langle YZ_*Z_2\rangle_{P_1}\langle YZ_1Z_2\rangle_{P_1}}+\frac{\langle Z_*Z_2Z_3\rangle_{P_1}^2}{\langle YZ_*Z_2\rangle_{P_1}\langle YZ_*Z_3\rangle_{P_1}\langle YZ_2Z_3\rangle_{P_1}}\\
&=\frac{A_{38}A_{47}}{X_{58}(A_{38}X_{58}-A_{47}X_{14})(A_{38}-X_{14})}-\frac{A_{38}A_{47}}{X_{14}(A_{38}X_{58}-A_{47}X_{14})(A_{47}-X_{58})},
\end{split}
\end{equation}
where we use \eqref{eq:coordinate8pt} and $A_{38}, A_{47}$ are given as \eqref{eq:a38a47}. We can check this corresponds to the canonical function of the $\mathcal{S}^{P_1}$.
We can also check that the terms $[Z_*,Z_1,Z_2]_{P_1}, [Z_*,Z_2,Z_3]_{P_1}$ are corresponding to the residues of the recursion relation $m^{{P_1}_L}_{34567I}\times m^{{P_1}_R}_{I812}, m^{{P_1}_L}_{456I}\times m^{{P_1}_R}_{I78123}$ respectively.
Next, we consider the $\mathcal{S}^{P_2}$. The constraints defining it inside the kinematic space are given in \eqref{eq:p4n8constraints}. We choose $\{X_{14},X_{16}\}$ as the basis and denote the vertices of this stokes polytope as
 \begin{equation}
 \begin{split}
&\{X_{14},X_{16}\},\{X_{14},X_{58}\},\{X_{58},X_{83}\},\{X_{83},X_{36}\},\{X_{36},X_{16}\} \\
&\rightarrow \mathcal{Z}^{P_2}_*, \mathcal{Z}^{P_2}_1, \mathcal{Z}^{P_2}_2, \mathcal{Z}^{P_2}_3,\mathcal{Z}^{P_2}_4.
\end{split}
\end{equation}  
Similarly we can triangulate this polytope as $(b)$ in Figure \ref{fig:stokespolytope}. Then the canonical function is given as
\begin{equation}
\begin{split}
&[Z_*,Z_1,Z_2]_{P_2}+[Z_*,Z_2,Z_3]_{P_2}+[Z_*,Z_3,Z_4]_{P_2}\\
&=\frac{A_{58}A_{38}}{X_{14}(A_{38}X_{16}-A_{58}X_{14})(A_{58}-X_{16})}+\frac{A_{36}A_{38}^2}{(X_{14}-A_{38})(A_{38}X_{16}-A_{58}X_{14})(A_{36}X_{14}-A_{38}X_{14}+A_{38}X_{16})}\\
&+\frac{(A_{36}-A_{58})(A_{36}+A_{38}-A_{58})}{X_{16}(A_{36}+A_{38}-A_{58}-X_{14}+X_{16})(A_{36}X_{14}-A_{38}X_{14}+A_{38}X_{16})}
\end{split}
\end{equation}
where $A_{58}, A_{36}$ are given as \eqref{eq:a58a36}. We can also check that these three terms are corresponding to the residues of the recursion relation $m^{{P_2}_L}_{567I}\times m^{{P_2}_R}_{I81234}, m^{{P_2}_L}_{34567I}\times m^{{P_2}_R}_{I812}, m^{{P_2}_L}_{345I}\times m^{{P_2}_R}_{I67812}$ respectively.
The sum of these functions with appropriate weight corresponds to the eight-point amplitude:
\begin{equation}
M_{p=4,n=8}=\alpha_{P_1}\sum_{\sigma}\sum_{i=1}^2[Z_*,Z_i,Z_{i+1}]_{{P_1},\sigma}+\alpha_{P_2}\sum_{\sigma'}\sum_{i=1}^3[Z_*,Z_i,Z_{i+1}]_{{P_2},\sigma'},
\end{equation}
where $\sigma,\sigma'$ range over all the cyclic permutations and $\alpha_{P_1}=\frac{1}{3}$ and $\alpha_{P_2}=\frac{1}{6}$. \par
\begin{figure}[t]
\begin{center}
\includegraphics[clip,width=17.0cm]{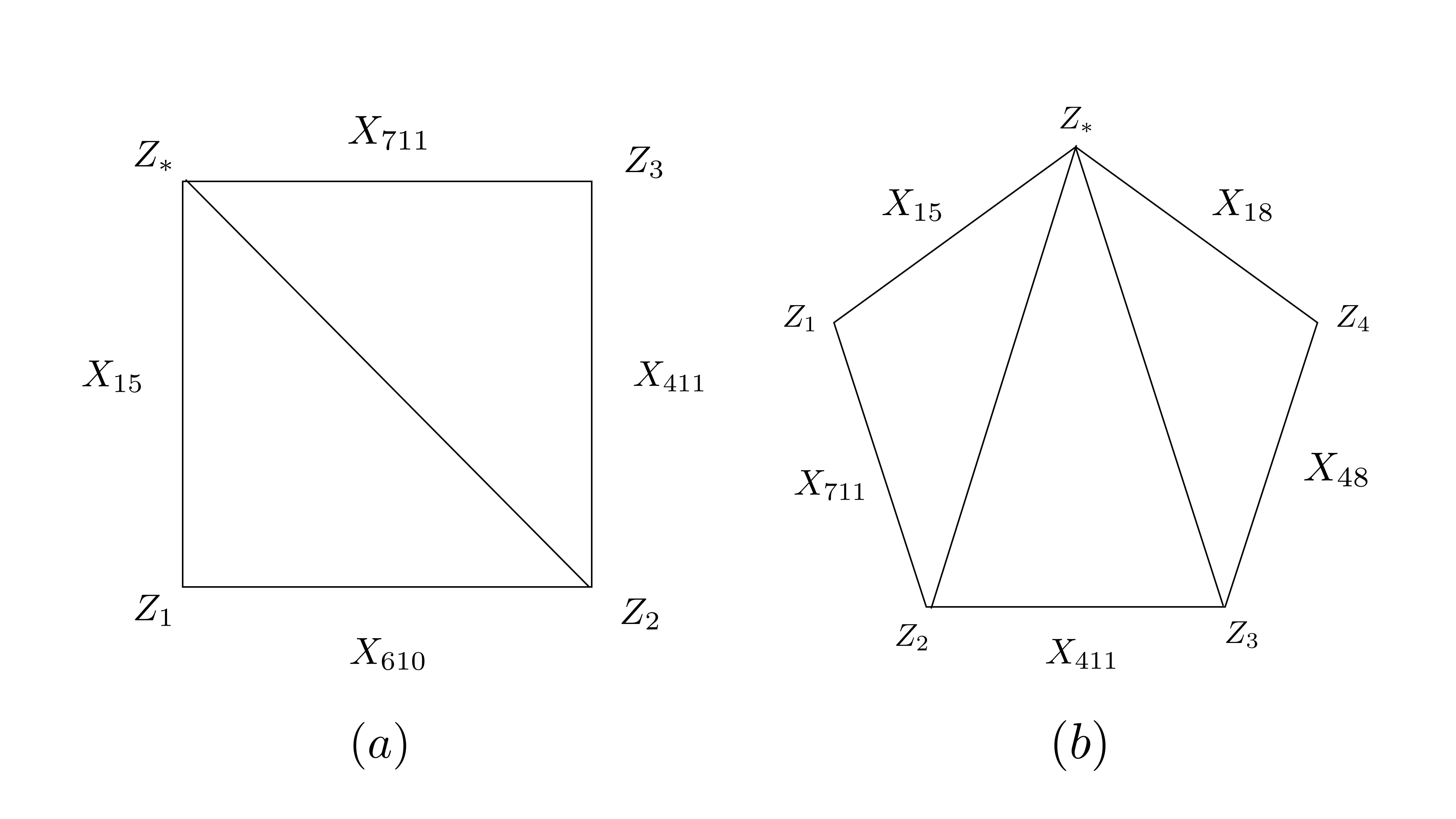}
 \caption{Triangulation of $p=5, n=11$ accordiohedra.}
 \label{fig:p5accordiohedron}
\end{center}
  \end{figure}

This representation is familiar from the BCFW representation of tree NMHV amplitudes for $\mathcal{N}=4$ SYM or $\phi^3$ tree amplitudes. %The important difference is the sum over all primitive graphs with appropriate weight. 
Similarly, we can obtain this representation for more higher point case. In the $n=10$ case, the stokes polytope becomes three-dimensional polytope  and we can triangulate explicitly. We put the results in the appendix.\\ \\
$\bold{p=5, n=11}$ case:  There are two primitive $p$-angulations $P_1=(15,711),P_2=(15,18)$. Two $Q$-compatible sets and the constraints are given as \eqref{eq:p5n11Qsets}, \eqref{eq:p5n11constraints}. The $Q_5(P_1)$ accordiohedron is a two-dimensional square. We can triangulate the accordiohedron as $(a)$ in Figure \ref{fig:p5accordiohedron} and the canonical function is given as
\begin{equation}
\begin{split}
\label{eq:p5n11Qtriangulation}
&[Z_*,Z_1,Z_2]_{P_1}+[Z_*,Z_2,Z_3]_{P_1}\\
&=\frac{A_{411}A_{610}}{X_{15}(X_{15}+X_{711}-A_{610})(A_{610}X_{15}-A_{411}X_{15}-A_{411}X_{711})}\\
&+\frac{A_{411}(A_{411}-A_{610})}{(A_{411}-X_{15})X_{711}(-A_{610}X_{15}+A_{411}X_{15}+A_{411}X_{711})}
\end{split}
\end{equation}
where $A_{411},A_{610}$ are given as \eqref{eq:a411a610}.  
  We can see that these two terms are corresponding to the residues at $\hat{X}_{411}, \hat{X}_{610}=0$ respectively. Similarly, we can compute the canonical function from the triangulation of the $Q_5(P_2)$ accordiohedron which is given as $(b)$ in Figure \ref{fig:p5accordiohedron}. 
\begin{equation}
\begin{split}
\label{eq:p5n11Q2triangulation}
&[Z_*,Z_1,Z_2]_{P_2}+[Z_*,Z_2,Z_3]_{P_2}+[Z_*,Z_3,Z_4]_{P_2}\\
&=\frac{A_{411}A_{711}}{X_{15}(X_{18}-A_{711})(A_{711}X_{15}-A_{411}X_{18})}\\
&+\frac{A_{411}^2(A_{711}+A_{48}-A_{411})}{(A_{411}-X_{15})(X_{15}A_{411}-X_{18}A_{411}-A_{48}X_{15})(-A_{711}X_{15}+A_{411}X_{18})}\\
&+\frac{A_{48}(A_{48}-A_{411})}{X_{18}(X_{15}A_{411}-X_{18}A_{411}-A_{48}X_{15})(A_{48}-X_{15}+X_{18})},
\end{split}
\end{equation}
where $A_{48},A_{711}$ are given as \eqref{eq:a48a711}. These three terms are corresponding to the residues at $\hat{X}_{48}, \hat{X}_{411}, \hat{X}_{711}=0$ respectively. Then the 11-point amplitude is given as
\begin{equation}
M_{p=5,n=11}=\alpha^{P_1}_{5,11}\sum_{\sigma}\sum_{i=1}^2[Z_*,Z_i,Z_{i+1}]_{{P_1},\sigma}+\alpha^{P_2}_{5,11}\sum_{\sigma'}\sum_{i=1}^3[Z_*,Z_i,Z_{i+1}]_{{P_2},\sigma'},
\end{equation}
where $\alpha^{P_1}_{5,11}=\frac{3}{11}$ and $\alpha^{P_2}_{5,11}=\frac{2}{11}$.\\ \\
%---------------------------------------------------------------------------------
$\bold{p=6, n=14}$ case: There are three $Q$-compatible sets $Q_6(P_1),Q_6(P_2),Q_6(P_3)$ which given in \eqref{eq:qsetn14p6}. 
We can compute the canonical function from the triangulation and the 14-point amplitude is given as
\begin{equation}
\begin{split}
M_{p=6,n=14}&=\alpha^{P_1}_{6,14}\sum_{\sigma}\sum_{i=1}^2[Z_*,Z_i,Z_{i+1}]_{{P_1},\sigma}+\alpha^{P_2}_{6,14}\sum_{\sigma}\sum_{i=1}^2[Z_*,Z_i,Z_{i+1}]_{{P_2},\sigma'}\\
&+\alpha^{P_3}_{6,14}\sum_{\sigma''}\sum_{i=1}^3[Z_*,Z_i,Z_{i+1}]_{{P_3},\sigma''},
\end{split}
\end{equation}
where $\alpha^{P_1}_{6,14}=\frac{1}{3}$, $\alpha^{P_2}_{6,14}=\frac{1}{6}$ and $\alpha^{P_3}_{6,14}=\frac{1}{3}$. The explicit representation for each team is written in the appendix.
%----------------------------------------------------------------------------------
\subsection{General formula for $\phi^p$ tree amplitudes}
The triangulation of the general associahedron is given in \cite{He:2018svj}. Since the accordiohedron is the simple polytope which is same as the associahedron, we can triangulate it by applying the same formula of \cite{He:2018svj}. The triangulation of the $D_p$ dimensional accordiohedron $\mathcal{AC}^{P}_{p,D_p}$ is given as
\begin{equation}
\sum_{i_1,j_1}\sum_{i_2,j_2}\cdots \sum_{i_{D_p},j_{D_p}}[Z^{(0)},Z_{i_1,j_1}^{(1)},\cdots,Z_{i_{D_p-1},j_{D_p-1}}^{(D_p-1)},Z_{i_{D_p},j_{D_p}}^{(D_p)}]_{P}
\end{equation}
Here we briefly see how to obtain this formula following with \cite{He:2018svj}. First we choose an origin $Z^{(0)}$ and connect to all co-dimension one facets that are not adjacent to $Z^{(0)}$. We denote these facets as $X_{i_1,j_1}$. Next, for each facet $X_{i_1,j_1}$, we triangulate it by connecting an origin $Z_{i_1,j_1}^{(1)}$ to all of its facets $X_{i_2,j_2}$. Continuing this for $D_p$ step and sum over all pairs $(i_1,j_1),(i_2,j_2),\dots$, we can triangulate this polytope.\par
By using this formula, the general $n$-point amplitude for $\phi^p$ is given as
\begin{equation}
\label{eq:generalformulaphip}
M_{p,n}=\sum_{a=1}^{p_n}\sum_{\sigma}\sum_{i_1,j_1}\cdots \sum_{i_{D_p},j_{D_p}}\alpha_{p,n}^{P_a}[Z^{(0)},Z_{i_1,j_1}^{(1)},\cdots,Z_{i_{D_p-1},j_{D_p-1}}^{(D_p-1)},Z_{i_{D_p},j_{D_p}}^{(D_p)}]_{P_a,\sigma}.
\end{equation}
where $p_n$ is a number of the primitive $p$-angulation and $\alpha_{p,n}^{P_a}$ is the weight.\par
The number $p_n$ for $\phi^p$ case is obtained in \cite{Raman:2019utu}. The explicit formula for this number is given as
\begin{equation}
p_n=\begin{cases}
    \frac{1}{n}F_{D_p,n}+\frac{1}{2}F_{(D_p+1)/2,p}+\frac{1}{n}\sum_{d=\text{Gcd}(D_p,p)}\phi(p)\frac{1}{d} 
    \begin{pmatrix}
      (p-1)D_p+\frac{p}{d} \\
      D_p 
    \end{pmatrix}
  & (\text{if $n$ is odd}) \\
     \frac{1}{n}F_{D_p,n}+\frac{1}{n}\sum_{d=\text{Gcd}(D_p,p)}\phi(p)\frac{1}{d} 
    \begin{pmatrix}
      (p-1)D_p+\frac{p}{d} \\
      D_p 
    \end{pmatrix}
& (\text{if $n$ is even})
  \end{cases}.
\end{equation}
where $F_{D_p,n}$ is the Fuss Catalan number 
\begin{equation}
F_{D_p,n}=\frac{1}{n(n-p)+p} \begin{pmatrix}
      (p-1)D_p \\
      D_p 
    \end{pmatrix}.
\end{equation}
And $\phi(d)$ is given by:
\begin{equation}
\phi(d)=d\left(1-\frac{1}{p_1}\right)\left(1-\frac{1}{p_2}\right)\cdots\left(1-\frac{1}{p_r}\right)
\end{equation}
where each $p_i$ comes from the prime factorization of $d$:
\begin{equation}
d=p_a^{a_1}p_2^{a_2}\cdots p_r^{a_r},\ \ p_i\ \text{are the prime numbers.}
\end{equation}
As we have seen in section \ref{sec:weightfactorization}, these weights $\alpha_{P_a}$ are determined from the factorization of the accordiohedron.
%---------------------------------------------------------------------------------------------------------------------
\section{Recursion from the projective triangulation}
\label{sec:projectiverecursion}
In section \ref{sec:recursion relation}, we introduced the ``BCFW"-like recursion relation for the $\phi^p$ amplitudes. This recursion is interpreted as a triangulation of accordiohedra. Recently, in  \cite{Arkani-Hamed:2019vag} a new triangulation called ``projective triangulation" for generalized ABHY-associahedra was introduced. From this new triangulation, we can obtain a new recursion relation called ``projective recursion relation". The field-theoretical derivation of the projective recursion was obtained in \cite{Yang:2019esm}. In this section, we apply this new recursion for $\phi^p$ amplitudes and interpret this as a projective triangulation of the accordiohedra.
%---------------------------------------------------------------------------------------------------------------------
\subsection{Projective recursion relation}
Here we briefly review the projective recursion relation by following \cite{Yang:2019esm}. First we consider a rescaling of $k$ variables
\begin{equation}
X_{A_i}\rightarrow zX_{A_i},\ \ i=1,2,\dots,k,
\end{equation}
here we choose $X_{A_i}$ as part of the basis of planar variables and $1\leq k\leq D_p$. Similar to the BCFW-like recursion, we consider the integral
 \begin{equation}
 \label{eq:integrand}
\oint \frac{z^k}{z-1}m_{p,n}^{P}(zX,C),
\end{equation}
where $m_{p,n}^{P}(zX,C)$ is the shifted $n$-point amplitude. By repeating same procedure of section \ref{sec:recursion relation}, we obtain the recursion relation formula
\begin{equation}
\label{eq:projectiverecursion}
m_{p,n}^{P}(X,C)=\sum_{B_i}\frac{z_{B_i}^k}{X_{B_i}}m^{P_L}_{a,\cdots,b-1,I}(z_{B_i}X,C)m^{P_R}_{I,b,\cdots,a-1}(z_{B_i}X,C),
\end{equation}
where the sum of $B_i$ runs over all the shifted planar variables. Here we denote the planar variables which depend on basis variables $X_{A_i}$ as 
\begin{equation}
X_{B_i}=C_{B_i}+X_i+\sum_{j=1}^k\lambda_{i,j}X_{A_j},
\end{equation}
where $\lambda_{i,j}$ are real numbers, $C_{B_i}$ and $X_i$ are linear combinations of constants $C$s and undeformed variable $X_{A_j}, j=k+1,\dots,D_p$. When $k=D_p$, this corresponds to \eqref{eq:BCFWrecursion}. Then we can interpret this as a generalization of the BCFW-like recursion relation.\par
In this derivation, we use the fact that the function $\frac{z^k}{z-1}m_{p,n}^{P}(zX,C)$ has no pole at $z=0$ and infinity. We prove these following with \cite{Yang:2019esm}. \\ \\
{\bf Proof of no pole at} $z=0$\par
First, we consider the pole at $z=0$. The canonical from of simple polytopes is given in \cite{Arkani-Hamed:2017tmz}:
\begin{equation}
\label{eq:canonicalformsimplepolytope}
\Omega=\sum_{v\in \text{vertices}}\prod_{\text{facet} f \in v} \frac{1}{X_f},
\end{equation}
where the facets $f$ are determined from $X_f=0$. Since a $d$-dimensional simple polytope has $d$ facets adjacent to any vertex $v$, the order of the product of $\frac{1}{X_f}$ becomes as
\begin{equation}
\label{eq:product}
\prod_{\text{facet} f \in v}\frac{1}{X_f}=O\left(\frac{1}{X^d}\right).
\end{equation}
If we rescale $k$ basis variables as
\begin{equation}
\label{eq:rescalekvariables}
X_{A_i}\rightarrow zX_{A_i},\ \ i=1,\dots,k,
\end{equation}
at most $k$ of the $d$ $X_{ij}$'s that appear in the product \eqref{eq:product} have a $z$ dependence. Then the product \eqref{eq:product} can at most have an order $k$ pole at $z=0$, and this is cancelled by the $z^k$ factor of the numerator of the function $\frac{z^k}{z-1}m_{p,n}^{P}(zX,C)$. This is what we want to prove.
\\ \\
{\bf Proof of no pole at infinity}\par
Here we rescale $k$ basis variables \eqref{eq:rescalekvariables} and consider the behavior of function $\frac{z^k}{z-1}m_{p,n}^{P}(zX,C)$ at $z\rightarrow \infty$. To see this, we need to consider the $O(\frac{1}{z^k})$ contribution of the canonical function \eqref{eq:canonicalformsimplepolytope}. This can be obtained that $k$ of the basis variables $X_{A_i}$ in the denominator of \eqref{eq:rescalekvariables} shifted and $(D_p-k)$ of them unshifted. Here we denote the unshifted variables as $\overline{X}_j$ and group the terms of \eqref{eq:canonicalformsimplepolytope} having the same $(D_p-k)$ $\overline{X}_j$'s;
\begin{equation}
\left(\sum_{\substack{\text{vertices shared by} \\D_p-k\ \overline{X}_j}}\frac{1}{\prod^kX}\right)\frac{1}{\prod^{D_p-k}_{j=1}\overline{X}_{j}}.
\end{equation}
Only the quantity in the bracket is affected by the rescaling.
When all these $\overline{X}_{j}\rightarrow0$, this quantity is the canonical function of the polytope which is the intersection of all these facets $\overline{X}_{j}$ and we denote as $m_{p,k}^{P}(X,C)$. When $z\rightarrow \infty$, 
\begin{equation}
\lim_{z\rightarrow \infty} m_{p,k}^{P}(X,C)\sim\frac{1}{z^k}m_{p,k}^{P}(X,0)\rightarrow0.
\end{equation}
Here we used the soft condition of the $m_{p,k}^{P}(X,C)$. We can easily verify this similarly as \eqref{eq:softfact}. Then the function $\frac{z^k}{z-1}m_{p,n}^{P}(zX,C)$ has no pole at infinity.\par
In \cite{Yang:2019esm}, it was proven that this recursion relation \eqref{eq:projectiverecursion} for the ABHY polytope is equivalent to the ``projective triangulation". We will see that in the accordiohedron case, this recursion relation can be interpreted as the projective triangulation of the accordiohedron.
%---------------------------------------------------------------------------------------------------------------------
\subsection{One-variable projective recursion}
In this section, we consider a interesting case: one-variable rescaling 
\begin{equation}
\label{eq:onevariableshift}
X_A\rightarrow zX_A.
\end{equation}
Under the rescaling \eqref{eq:onevariableshift}, the shift parameter is determined by solving the equation
\begin{equation}
\hat{X}_{B_i}=C_{B_i}+X_i+\sum_{j=1}^{k-1}\lambda_{i,j}X_{A_j}+z_{B_i}\lambda_{i,k}X_{A_k}=0.
\end{equation}
The result is
\begin{equation}
z_{B_i}=-\frac{C_{B_i}+X_i+\sum_{j=1}^{k-1}\lambda_{i,j}X_{A_j}}{\lambda_i Z_A}=1-\frac{X_{B_i}}{\lambda_{i,k} X_{A_k}}.
\end{equation}
Then the shifted basis variables are written as 
\begin{equation}
\label{eq:onevariablesubtitute}
X_A \rightarrow z_{B_i}X_A=X_A-\frac{1}{\lambda_{i,k}}X_{B_i}.
\end{equation}
When we determine $z_{B_i}$ from the equation $\hat{X}_{B_i}=0$, the another variable $X_C=C_{C}+\lambda X_A+\sum_{j=1}^{D_p-1}\lambda
_{ij}X_{A_j}$ which depends on the shifted basis variable becomes as 
\begin{equation}
\begin{split}
\label{eq:onevariablesubtitute2}
\hat{X}_{C_i}&=C_{C_i}+X_{i}+\sum_{j=1}^{k-1}\lambda_{i,j}X_{A_j}+z_{B_i}\lambda_{i,k}X_{A_k}=X_{C_i}-X_{B_i}.
\end{split}
\end{equation}
From these results and
\begin{equation}
\frac{z_{B_i}}{X_{B_i}}=\left(\frac{1}{X_{B_i}}-\frac{1}{\lambda_iX_{A}}\right),
\end{equation}
we can obtain the one-variable rescaling recursion
\begin{equation}
\label{eq:onevariablerecursionformula}
m_{p,n}^{P}(X,C)=\sum_{B_i}\left(\frac{1}{X_{B_i}}-\frac{1}{\lambda_iX_{A}}\right)\hat{m}^{P_L}_{a,\cdots,b-1,I}\times\hat{m}^{P_R}_{I,b,\cdots,a-1},
\end{equation}
where $\hat{m}$ means that before sum over $B_i$, we need to make the replacement \eqref{eq:onevariablesubtitute} and \eqref{eq:onevariablesubtitute2}. This recursion relation was also derived from the general properties of canonical forms of simple polytopes \cite{Salvatori:2019phs}.
Next, we see some examples of this one-variable projective recursion.\\ \\
$\bold{p=4, n=8}$ case:\\
Let's consider a one-variable rescaling 
\begin{equation}
X_{14}\rightarrow zX_{14}.
\end{equation}
In this case, there are two $Q$-compatible sets \eqref{eq:Qcompatible8pt} and \eqref{eq:Q2compatible8pt}. In the $Q(P_1)$ set, only $X_{38}$ will be shifted. From \eqref{eq:onevariablerecursionformula},
\begin{equation}
\begin{split}
m_{p=4,n=8}^{P_1}(X,C)&=\left(\frac{1}{X_{38}}+\frac{1}{X_{14}}\right)\hat{m}^{{P_1}_L}_{34567I}\times\hat{m}^{{P_1}_R}_{I812}\\
&=\left(\frac{1}{X_{38}}+\frac{1}{X_{14}}\right)\left(\frac{1}{X_{58}}+\frac{1}{X_{47}}\right).
\end{split}
\end{equation}
This is corresponding to the $p=4, n=8, Q(P_1)$-compatible amplitude. This is the canonical function of the rectangle form by $X_{38},X_{58}$ as its edges. We can interpret this result as the prism formed by projecting the facet $X_{38}$ onto line $X_{14}$. \par
\begin{figure}[t]
\begin{center}
\includegraphics[clip,width=17.0cm]{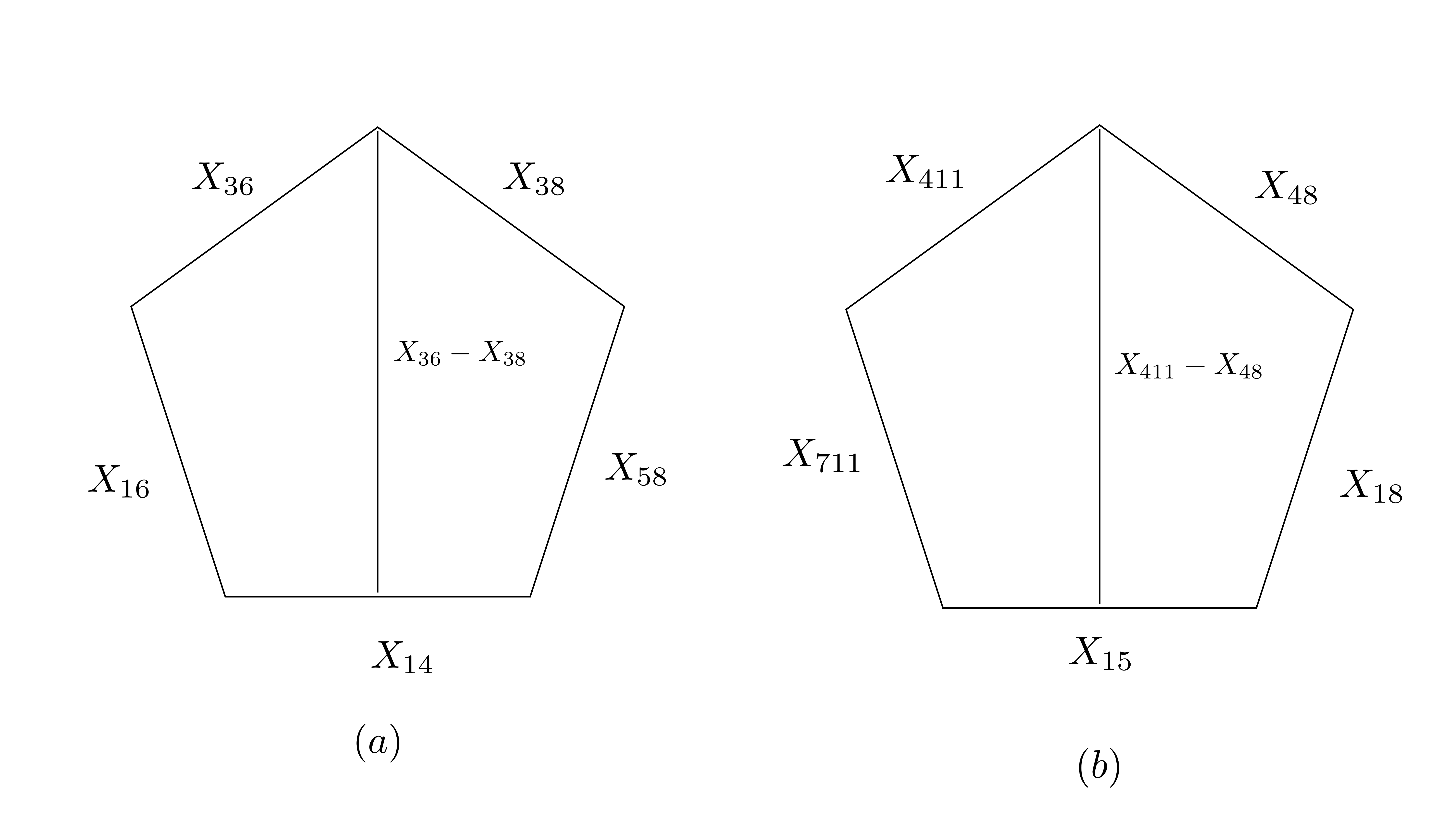}
 \caption{Projecting triangulation for $p=4, n=8$ and $p=5, n=11$.}
 \label{fig:pentagon5pt}
\end{center}
  \end{figure}

 
In the $Q(P_2)$ set, $X_{38}$ and $X_{36}$ will be shifted. Then from the recursion formula, 
\begin{equation}
\begin{split}
m_{4,8}^{P_2}(X,C)&=\left(\frac{1}{X_{38}}+\frac{1}{X_{14}}\right)\hat{m}^{{P_2}_L}_{34567I}\times\hat{m}^{{P_2}_R}_{I812}+\left(\frac{1}{X_{36}}+\frac{1}{X_{14}}\right)\hat{m}^{{P_2}_L}_{345I}\times\hat{m}^{{P_2}_R}_{I67812}\\
&=\left(\frac{1}{X_{38}}+\frac{1}{X_{14}}\right)\left(\frac{1}{X_{36}-X_{38}}+\frac{1}{X_{58}}\right)+\left(\frac{1}{X_{36}}+\frac{1}{X_{14}}\right)\left(\frac{1}{X_{38}-X_{36}}+\frac{1}{X_{16}}\right).
\end{split}
\end{equation}
It is easy to check that this corresponds to the $p=4, n=8, Q(P_2)$-compatible amplitude. These two terms are the canonical function of the two 2-dimensional prisms as $(a)$ in Figure \ref{fig:pentagon5pt}. 
\\ \\
$\bold{p=4, n=10}$ case:\\
In this case, there are 7 primitive $\{P_1,P_2,\dots,P_7\}$. Here we consider $P_1=(14,510,69)$, the $Q$-compatible set is given as \eqref{eq:qsetn10}. Let's consider a one-variable rescaling 
\begin{equation}
X_{14}\rightarrow zX_{14}.
\end{equation}
Only $X_{310}$ will be shifted and the recursion formula becomes as
\begin{equation}
\begin{split}
m_{p=4,n=10}^{P_1}(X,C)&=\left(\frac{1}{X_{14}}+\frac{1}{X_{310}}\right)\hat{m}^{{P_1}_L}_{3456789I}\times\hat{m}^{{P_1}_R}_{I1012}\\
&=\left(\frac{1}{X_{14}}+\frac{1}{X_{310}}\right)\left(\frac{1}{X_{510}X_{58}}+\frac{1}{X_{49}X_{58}}+\frac{1}{X_{510}X_{69}}+\frac{1}{X_{49}X_{69}}\right).
\end{split}
\end{equation}
This corresponds to the $p=4, n=10, Q(P_1)$-compatible amplitude. This is the product of the canonical functions of the line $\{X_{14}, X_{310}\}$ and the rectangle form by $\{X_{58},X_{510}, X_{69}, X_{49}\}$. We can interpret this result as the prism formed by projecting the facet $X_{310}$ onto line $X_{14}$.\\ \\
$\bold{p=5, n=11}$ case:\\
Let's consider a one-variable rescaling 
\begin{equation}
X_{15}\rightarrow zX_{15}.
\end{equation}
In this case, there are two $Q$-compatible sets \eqref{eq:p5n11Qsets}. In the $Q(P_1)$ set, only $X_{411}$ will be shifted and the recursion formula becomes as
\begin{equation}
\begin{split}
m_{p=5,n=11}^{P_1}(X,C)&=\left(\frac{1}{X_{411}}+\frac{1}{X_{15}}\right)\hat{m}^{{P_1}_L}_{45678910I}\times\hat{m}^{{P_1}_R}_{I11123}\\
&=\left(\frac{1}{X_{411}}+\frac{1}{X_{15}}\right)\left(\frac{1}{X_{117}}+\frac{1}{X_{610}}\right).
\end{split}
\end{equation}
This corresponds to the $p=5, n=11, Q(P_1)$-compatible amplitude. This is the canonical function of the the prism formed by projecting the facet $X_{411}$ onto line $X_{15}$. Similarly, in the $Q(P_2)$ set $X_{48}$ and $X_{411}$ will be shifted. Then from the recursion formula, 
\begin{equation}
\begin{split}
m_{5,11}^{P_2}(X,C)&=\left(\frac{1}{X_{48}}+\frac{1}{X_{15}}\right)\hat{m}^{{P_2}_L}_{4567I}\times\hat{m}^{{P_2}_R}_{I891011123}+\left(\frac{1}{X_{411}}+\frac{1}{X_{15}}\right)\hat{m}^{{P_2}_L}_{45678910I}\times\hat{m}^{{P_2}_R}_{I11123}\\
&=\left(\frac{1}{X_{48}}+\frac{1}{X_{15}}\right)\left(\frac{1}{X_{411}-X_{48}}+\frac{1}{X_{18}}\right)+\left(\frac{1}{X_{411}}+\frac{1}{X_{15}}\right)\left(\frac{1}{X_{48}-X_{411}}+\frac{1}{X_{711}}\right).
\end{split}
\end{equation}
This corresponds to the $p=5, n=11, Q(P_2)$-compatible amplitude. We can also see that these two terms are the canonical functions of the two 2-dimensional prisms as $(b)$ in Figure \ref{fig:pentagon5pt}.
%---------------------------------------------------------------------------------------------------------------------
\section{Discussions}
In this paper, we have investigated the weights of the accordiohedron and the one-parameter recursion relations of the $\phi^p$ tree amplitude. The main difference with the $\phi^3$ case is that there is no single polytope which represents complete scattering amplitude. For each $Q$-compatible set of graphs of $\phi^p$, we can define the accordiohedron as the positive geometry and the full scattering amplitudes are given as a weighted sum over all of these accordiohedra. \par
We have determined these weights from the factorization property of the accordiohedra. This means that even in this $\phi^p$ case, the geometry of the accordiohedron is enough to calculate scattering amplitudes. \par
We also have constructed one-parameter recursion relations of the $\phi^p$ tree amplitudes. In the case of $\phi^p$, we need to consider these recursion relations for each $Q$-compatible set of graphs. After some examples of the BCFW-like recursion, we have obtained the all-multiplicity result for the $\phi^p$ tree amplitudes. In addition to this, we constructed the projective recursion relation of the $\phi^p$ tree amplitudes. We calculated some examples of the one-variable projective recursion. This can be interpreted as the projective triangulation of the accordiohedron.\par
There are many open questions for future studies. One of the issues is to consider the polynomial interaction case. In the case of $\lambda_3\phi^3+\lambda_4\phi^4$, it is known that the positive geometry is the accordiohedron \cite{Jagadale:2019byr}. Even in this case, there are some accordiohedra for each amplitude and we need to determine the weights to obtain the amplitude. Then is it possible to determine the weights from the factorization of the accordiohedra for this polynomial interaction case? Another question is whether is it possible to apply the one-parameter recursions to this polynomial case. We plan to address these questions in the future.

%---------------------------------------------------------------------------------------------------------------------
\section*{Acknowledgements}
We would like to thank Song He for suggesting the problem of the weights and projective recursions. We would also like to thank Prashanth Raman for advice and comments on improving the draft.

\appendix
\section{Some details for $p=4, n=10$ case}
{\bf Triangulation of the stokes polytope}\\\\
There are 7 primitive quadrangulation
\begin{equation}
\begin{split}
P_1&=(14,510,69),\ P_2=(14,16,18),\ P_3=(14,16,69),\ P_4=(14,49,69),\\
P_5&=(14,47,710),\ P_6=(14,510,710),\ P_7=(14,16,710). 
\end{split}
\end{equation}
First, we consider $P_1$, the $Q$-compatible set is given as \eqref{eq:qsetn10}.
The stokes polytope for this case is 8-vertices three dimensional polytope. The constraints defining it in the kinematic space are given as
\begin{equation}
\begin{split}
s_{ij}&=-C_{ij},\ \text{for}\ i\leq i<j\leq9\ \text{with} |i-j| \geq2\\
X_{13}&=d_{13}, X_{410}=d_{410}, X_{59}=d_{59}, X_{68}=d_{68}.
\end{split}
\end{equation}  
We label all of the vertices by the three adjacent facets as
\begin{equation}
\begin{split}
&Z^{(0)}=\{X_{14},X_{510},X_{69}\}, Z_1=\{X_{58},X_{49},X_{310}\},  Z_2=\{X_{58},X_{510},X_{310}\}, Z_3=\{X_{69},X_{510},X_{310}\}, \\
&Z_4=\{X_{49},X_{14},X_{58}\}, Z^{(1)}_{310}=\{X_{69},X_{310},X_{49}\},Z^{(1)}_{49}=\{X_{49},X_{69},X_{14}\},Z^{(1)}_{58}=\{X_{58},X_{14},X_{510}\}.
\end{split}
\end{equation}  
The direct computation gives 
\begin{equation}
\begin{split}
&Z^{(0)}=(1,0,0,0), Z_1=(1,A_{310},A_{49},A_{58}), Z_2=(1,A_{310},0,A_{58}), Z_3=(1,A_{310},0,0), \\
&Z_4=(1,0,A_{49},A_{58}),Z^{(1)}_{310}=(1,A_{310},A_{49},0),Z^{(1)}_{49}=(1,0,A_{49},0),Z^{(1)}_{58}=(1,0,0,A_{58}).
\end{split}
\end{equation}  
where
\begin{equation}
\begin{split}
A_{310}=\sum_{\substack{1\leq a <3 \\ 4\leq b<10}}C_{ab},\ A_{58}=d_{68}+d_{59}+C_{59},\ A_{49}=d_{59}+d_{410}+C_{410},
\end{split}
\end{equation}
and we omitted the label of ${Q(P_1)}$ ($Z^{P_1}\rightarrow Z$). From the general formula, the triangulation of this polytope is given as
\begin{equation}
\begin{split}
&[Z^{(0)},Z^{(1)}_{310},Z_1,Z_2]_{P_1}+[Z^{(0)},Z^{(1)}_{310},Z_2,Z_3]_{P_1}+[Z^{(0)},Z^{(1)}_{49},Z_4,Z_1]_{P_1}\\
&+[Z^{(0)},Z^{(1)}_{49},Z_1,Z^{(1)}_{310}]_{P_1}
+[Z^{(0)},Z^{(1)}_{58},Z_2,Z_1]_{P_1}+[Z^{(0)},Z^{(1)}_{58},Z_1,Z_4]_{P_1},
\end{split}
\end{equation}  
We can compute each term, for example:
\begin{equation}
\begin{split}
&[Z^{(0)},Z^{(1)}_{310},Z_1,Z_2]_{P_1}\\
&=\frac{A_{310}^3A_{49}A_{58}}{(A_{310}-X_{14})(A_{310}X_{510}-A_{49}X_{14})(A_{310}X_{69}-A_{58}X_{14})(A_{310}A_{49}X_{69}+A_{310}A_{58}X_{510}-A_{49}A_{58}X_{14})},
\end{split}
\end{equation}  
\begin{equation}
\begin{split}
&[Z^{(0)},Z^{(1)}_{310},Z_2,Z_3]_{P_1}\\
&=\frac{A_{310}A_{49}A_{58}}{(-A_{310}+X_{14})X_{510}X_{69}(A_{310}A_{49}X_{69}+A_{310}A_{58}X_{510}-A_{49}A_{58}X_{14})},
\end{split}
\end{equation}  
\begin{equation}
\begin{split}
&[Z^{(0)},Z^{(1)}_{49},Z_4,Z_1]_{P_1}\\
&=\frac{A_{310}A_{49}A_{58}}{X_{14}(A_{49}-X_{510})(A_{58}X_{14}-A_{310}X_{69})(A_{49}X_{69}-A_{58}X_{510})},
\end{split}
\end{equation}  
\begin{equation}
\begin{split}
&[Z^{(0)},Z^{(1)}_{49},Z_1,Z^{(1)}_{310}]_{P_1}\\
&=\frac{A_{310}A_{49}A_{58}}{X_{14}X_{69}(-A_{49}+X_{510})(A_{310}A_{49}X_{69}+A_{310}A_{58}X_{510}-A_{49}A_{58}X_{14})},
\end{split}
\end{equation}  
\begin{equation}
\begin{split}
&[Z^{(0)},Z^{(1)}_{58},Z_2,Z_1]_{P_1}\\
&=\frac{A_{310}A_{49}A_{58}}{X_{510}(-A_{49}X_{14}+A_{310}X_{510})(A_{58}-X_{69})(A_{58}X_{14}-A_{310}X_{69})},
\end{split}
\end{equation}  
\begin{equation}
\begin{split}
&[Z^{(0)},Z^{(1)}_{58},Z_1,Z_2]_{P_1}\\
&=\frac{A_{310}A_{49}A_{58}}{X_{14}(-A_{49}X_{14}+A_{310}X_{510})(A_{58}-X_{69})(-A_{58}X_{510}+A_{49}X_{69})}
\end{split}
\end{equation}  
Then the sum of these terms is
\begin{equation}
\begin{split}
&\frac{1}{X_{14}X_{49}X_{58}}+\frac{1}{X_{310}X_{49}X_{58}}+\frac{1}{X_{14}X_{510}X_{58}}+\frac{1}{X_{310}X_{510}X_{58}}\\
&+\frac{1}{X_{14}X_{49}X_{69}}+\frac{1}{X_{310}X_{49}X_{69}}+\frac{1}{X_{14}X_{510}X_{69}}+\frac{1}{X_{310}X_{510}X_{69}}.
\end{split}
\end{equation}  
This corresponds to the canonical function of the stokes polytope.
%---------------------------------------------------------------------------------------------------------------------
\section{Explicit results of the triangulation for $p=6,n=14$ case}
\begin{equation}
\begin{split}
&[Z_*,Z_1,Z_2]_{P_1}+[Z_*,Z_2,Z_3]_{P_1}\\
&=\frac{A_{514}A_{813}}{(A_{514}-X_{16})X_{914}(-X_{914}A_{514}+X_{16}A_{813})}+\frac{A_{514}A_{813}}{(X_{914}-A_{813})X_{16}(X_{16}A_{813}-X_{914}A_{514})},\\
\ \\
&[Z_*,Z_1,Z_2]_{P_2}+[Z_*,Z_2,Z_3]_{P_2}\\
&=\frac{A_{514}A_{712}}{(X_{914}-A_{712})X_{914}(X_{914}A_{514}-X_{16}A_{712})}+\frac{A_{514}A_{712}}{(X_{16}-A_{514})X_{16}(X_{16}A_{712}-X_{914}A_{514})},\\
\ \\
&[Z_*,Z_1,Z_2]_{P_3}+[Z_*,Z_2,Z_3]_{P_3}+[Z_*,Z_3,Z_4]_{P_3}\\
&=\frac{A_{510}(A_{510}-A_{514})}{X_{110}(A_{510}+X_{110}-X_{16})(A_{514}X_{110}-A_{514}X_{16}+A_{510}X_{16})}\\
&+\frac{A_{514}^2(A_{510}-A_{514}+A_{914})}{(A_{514}-X_{16})(A_{514}X_{110}-A_{914}X_{16})(A_{514}X_{16}-A_{514}X_{110}-A_{510}X_{16})}\\
&+\frac{A_{514}A_{914}}{(A_{914}-X_{110})X_{16}(X_{110}A_{514}-X_{16}A_{914})}
\end{split}
\end{equation}
where
\begin{equation}
\begin{split}
A_{514}&=\sum_{\substack{1\leq a<5\\6\leq b<14}}C_{ab},\ \ A_{510}=\sum_{\substack{1\leq a<5\\6\leq b<10}}C_{ab}\\
A_{813}&=d_{814}+d_{913}+C_{813},\ \ A_{712}=d_{812}+d_{713}+C_{712}.
\end{split}
\end{equation}
%\section{Recursion relation for $p=6$ amplitudes}
%$\bold{n=10, p=6}$\\
%$\bold{n=14, p=6}$
\bibliographystyle{jhep}
\bibliography{ref} 
\end{document}